\RequirePackage{fix-cm}
\documentclass[twocolumn]{svjour3}          % twocolumn
\smartqed  % flush right qed marks, e.g. at end of proof
\usepackage{amssymb}
\usepackage{siunitx}
\usepackage{natbib}        
\usepackage{amsmath}
\usepackage{graphicx}
\usepackage[T1]{fontenc}
\usepackage[utf8,latin1]{inputenc}
\usepackage{epsfig}
\usepackage{txfonts}
\usepackage{threeparttable}
\usepackage{url}
\usepackage{color}
\usepackage{xcolor}

\usepackage{bm}
\usepackage{sgl-commands}
\usepackage[normalem]{ulem}
\newcommand{\tdist}{D_{\Delta t}}
\newcommand{\kext}{\kappa_{\rm{ext}}}

\newcommand{\Hunit}{km s$^{-1}{\rm Mpc}^{-1}$}
\newcommand{\chapintro}{\cite{Saha:2024}}
\newcommand{\chapgal}{\cite{Shajib:2023review}}

\newcommand{\chapmicro}{\cite{Vernardos:2024}}
\newcommand{\chapsn}{\cite{Suyu:2024}}
\newcommand{\chapsearch}{\cite{Lemon:2023}}

 \journalname{my journal}

\begin{document}

%\title{Chapter 7: Time-Delay Cosmography:  Measuring the Hubble Constant and other cosmological parameters from strong-lens systems}

\title{Time-Delay Cosmography:  Measuring the Hubble Constant and other cosmological parameters with strong gravitational lensing}

%\subtitle{}

\titlerunning{Time-Delay Cosmography}        % if too long for running head

\author{S.~Birrer$^{1, 2}\dagger$\thanks{$\dagger$ \email{simon.birrer@stonybrook.edu}}, M.~Millon$^{2, 3}$, D.~Sluse$^{4}$, A.~J.~Shajib$^{5, 6, 7}$, F.~Courbin$^{3, 8, 9}\ddagger$\thanks{$\ddagger$ \email{frederic.courbin@epfl.ch}}, 
S.~Erickson$^{2}$,
L.~V.~E.~Koopmans$^{10}$, S.~H.~Suyu$^{11, 12, 13}$, T.~Treu$^{14}$
% (additional authors)
}

\authorrunning{Birrer et al.} % if too long for running head

\institute{
$^{1}$ Department of Physics and Astronomy, Stony Brook University, Stony Brook, NY 11794, USA\\
$^{2}$ Kavli Institute for Particle Astrophysics and Cosmology and Department of Physics, Stanford University,
Stanford, CA 94305, USA\\
$^{3}$ Institute of Physics, Laboratory of Astrophysics, Ecole Polytechnique F\'ed\'erale de Lausanne (EPFL), Observatoire de Sauverny, 1290 Versoix, Switzerland\\
$^{4}$ STAR Institute, Quartier Agora - All\'ee du six Ao\^ut, 19c B-4000 Li\`ege, Belgium\\
$^{5}$ Department of Astronomy and Astrophysics, University of Chicago, Chicago, IL 60637, USA\\
$^{6}$ Kavli Institute for Cosmological Physics, University of Chicago, Chicago, IL 60637, USA\\
$^{7}$ NHFP Einstein Fellow\\
$^{8}$  ICC-UB Institut de Ci\`encies del Cosmos, University of Barcelona, Mart\'i Franqu\`es, 1, E-08028 Barcelona, Spain\\
$^{9}$  ICREA, Pg. Llu\'is Companys 23, Barcelona, E-08010, Spain\\
$^{10}$ Kapteyn Astronomical Institute, University of Groningen, P.O.Box 800, 9700AV Groningen, the Netherlands\\
$^{11}$ Max-Planck-Institut f{\"u}r Astrophysik, Karl-Schwarzschild-Str.~1, 85748 Garching, Germany\\
$^{12}$ Technical University of Munich, TUM School of Natural Sciences, Physics Department,  James-Franck-Stra\ss{}e~1, 85748 Garching, Germany\\
$^{13}$ Academia Sinica Institute of Astronomy and Astrophysics (ASIAA), 11F of ASMAB, No.1, Section 4, Roosevelt Road, Taipei 10617, Taiwan\\
$^{14}$ Department of Physics and Astronomy, University of California, Los Angeles CA 90095\\
}

%\institute{N. Name \at
%              Affiliation\\
%              Address \\
%		   City, Country\\
%              \email{name@somewhere.edu}           
% }

\date{Received: date / Accepted: date}

%\input{xxx}

\maketitle

\begin{abstract}
Multiply lensed images of a same source experience a relative time delay in the arrival of photons due to the path length difference and the different gravitational potentials the photons travel through. This effect can be used to measure absolute distances and the Hubble constant ($H_0$) and is known as time-delay cosmography.
The method is independent of the local distance ladder and early-universe physics and provides a precise and competitive measurement of $H_0$. With upcoming observatories, time-delay cosmography can provide a 1\% precision measurement of $H_0$ and can decisively shed light on the current reported 'Hubble tension'. 
This manuscript details the general methodology developed over the past decades in time-delay cosmography, discusses recent advances and results, and, foremost, provides a foundation and outlook for the next decade in providing accurate and ever more precise measurements with increased sample size and improved observational techniques.

% add further keywords with "\and" 
%\keywords{gravitational lensing: strong -- cosmology: Hubble constant} 
\end{abstract}

%\tableofcontents

%%%%%%%%
% This is an example for a chapter reference: \chapref{chapter1}.
% \\
% This is an example for a section reference: \secref{sec1:intuitive_teaser}.
% \\
% This is an example for a figure reference: \figref{fig1:Fig1}.
% \\
% This is an example for a table reference: \tabref{tab1:bursting_neutron}.
% \\
% This is an example for an equation reference: \eqref{eq1:opacity}.
%%%%%%%%

\label{chapter7}

\section{Introduction} \label{sec7:introduction}
The relative arrival times of multiply lensed sources can be used to measure an absolute distance of the Universe. The method, known to date as time-delay cosmography, was originally proposed over half a century ago, prior to the discovery of the first extragalactic gravitational lens, by \cite{Refsdal:1964}. Time-delay cosmography provides a one-step measurement of the Hubble constant ($H_0$), independent of the local distance ladder or probes anchored with sound horizon physics, such as the cosmic microwave background (CMB).

Almost a century after its first measurement, the Hubble constant $H_0$ still remains arguably the most debated number in cosmology. In the past few years, a tension has emerged between a number of local measurements, and inferences from early-Universe probes such as the cosmic microwave background (CMB) and Big Bang Nucleosynthesis, under the assumption of flat $\Lambda$ cold dark matter ($\Lambda$CDM) cosmology \citep[see, e.g.,][for recent summaries]{Verde:2019, Shah:2021, Abdalla22}. If this tension is real, and not due to unknown systematic uncertainties in multiple measurements and their analyses, it implies that the standard $\Lambda$CDM model is not sufficient and new physical ingredients beyond this model are required. From a theoretical standpoint, a number of possible solutions -- for example, involving early dark energy -- have been proposed \citep[e.g.,][and references therein]{KnoxMillea:2020, DiValentino:2021, Schoeneberg:2021}, often requiring fine-tuning of free parameters not to violate other observational constraints. From an observational standpoint, besides improving the precision of the measurements, significant attention has turned to the systematic investigation of unknown systematic uncertainties \citep[e.g.,][]{Riess:2019, Freedman:2019, Freedman:2020, Riess:2021, Mortsell:2021, Dainotti:2021, Riess23}.

This manuscript details the general methodology developed over the past decades in time-delay cosmography, discusses recent advances and results, and, foremost, provides a foundation and outlook for the next decade in providing accurate and ever more precise measurements with increased sample size and improved observational techniques.
This manuscript is becoming a chapter of a Space Science Reviews, Topical Collection "Strong Gravitational Lensing", eds. J. Wambsganss et al. 
We will refer throughout this manuscript to other chapters of the same collection covering a wide range of strong lensing theory and applications 
\citep[e.g.,][]{Saha:2024, Shajib:2023review, Suyu:2024, Vernardos:2024, Lemon:2023}. We refer to, e.g., \cite{TreuMarshall:2016, Suyu:2018} to provide more in-depth historical perspectives on the early years of the field, \cite{Treu:2022,T+S23} to a broader and less technical perspective on the opportunities of time-delay cosmography in this decade, and \cite{Moresco:2022} for a compact overview embedded within other cosmological probes.

We discuss primarily the methodology around lensed quasars as source objects to perform time-delay measurements because quasars are currently the most established sources with the most relevant current results. The use of lensed supernovae for cosmological and astrophysical studies is reviewed in detail by \citet{Suyu:2024} in the same series; interested readers can also find descriptions of lensed supernova cosmographic results until 2023 in \chapsn. For additional discussions on lensed supernovae and particularly on other types of lensed transients, such as gamma ray bursts and fast radio bursts, we refer to \citet{Oguri19} and \citet{Liao:2022}.
%We refer to  \citet{Oguri19} and \chapsn~using, for example, lensed supernovae or other transient phenomena that can be utilized to perform time-delay cosmographic measurements. 

This manuscript is organized as follows: In Section~\ref{sec7:distance_measure} we provide the general concept and physics to turn relative time delays between multiple images of the same source into distance measurements. Section~\ref{sec7:analysis_overview} provides an overview of the required analysis ingredients for individual lenses. Subsequent sections go into the details of these ingredients, the time-delay measurement (Section~\ref{sec7:time_delays}), determination of the lensing potential (Section~\ref{sec7:lens_potential}), and the study of the line of sight (LOS) of the lenses (Section~\ref{sec7:los}).
Section~\ref{sec7:cosmo_inferences} describes the cosmographic inferences and how to utilize a sample of lenses to perform an $H_0$ inference. 
In Section~\ref{secX:clusteres} we present an overview of the cosmographic method applied for galaxy clusters as the deflectors.
We summarize the current status and results obtained in Section~\ref{sec7:current_results}. Lastly, in Section~\ref{sec7:outlook}, we look in the future and discuss the potential and challenges lying ahead for the community.

\section{Time delays and the time-delay distance}\label{sec7:distance_measure}

\subsection{Lensing formalism and the Fermat Potential}
The deflection of light due to mass over- or under-density in the Universe lead to an angular displacement between the angle of the arriving photons, where we see the image, and the angle to the originating source ignoring lensing effects.

The lens equation formally describes the lensing distortions
\begin{equation}\label{eqn:lens_equation}
    \boldsymbol{\beta} = \boldsymbol{\theta} - \boldsymbol{\alpha}(\boldsymbol{\theta}),
\end{equation}
where $\boldsymbol{\beta}$ is the angular position of the source without the lensing effect, $\boldsymbol{\theta}$ is the corresponding angular coordinate on the sky as seen when lensed, and $\boldsymbol{\alpha}$ is the deflection angle that maps the image position to the source position in angular coordinates as seen from the observer.
We refer to e.g., \chapintro~ for further details into the theory, including that this formula is only valid for small angles.

There exists a scalar potential, the lensing potential $\phi$, such that the gradients correspond to the deflection field
\begin{equation}
    \nabla\phi(\boldsymbol{\theta}) = \boldsymbol{\alpha}(\boldsymbol{\theta}).
\end{equation}

The convergence of the potential $\phi$, $\kappa$, is half the Laplacian
\begin{equation}
    \kappa ({\boldsymbol{\theta }})={\frac {1}{2}}\nabla ^{2}\phi ({\boldsymbol{\theta }})
\end{equation}
and is given in the thin lens approximation for small angles as

\begin{equation} \label{eqn:kappa_sigma}
     \kappa ({\boldsymbol{\theta }})={\frac {\Sigma ({\boldsymbol{\theta }})}{\Sigma _{\rm crit}}}
\end{equation}
with $\Sigma ({\boldsymbol{\theta }})$ as the projected mass over- or under-density with respect to the mean background density and $\Sigma _{\rm crit}$  the critical surface density\footnote{not to be confused with the critical density of the universe}
\begin{equation} \label{eqn:sigma_crit}
    \Sigma _{\rm crit}={\frac {c^{2}D_{\rm s}}{4\pi G D_{\rm ds}D_{\rm d}}},
\end{equation}
where $c$ is the speed of light and $G$ the gravitational constant. $D_{\rm d}$ is the angular diameter distance to the lens, $D_{\rm s}$ is the angular diameter distance to the source, and $D_{\rm ds}$ is the angular diameter distance between the lens and the source, respectively. 

\begin{figure*}[htbp]
\centering
    \includegraphics[width=0.8\textwidth]{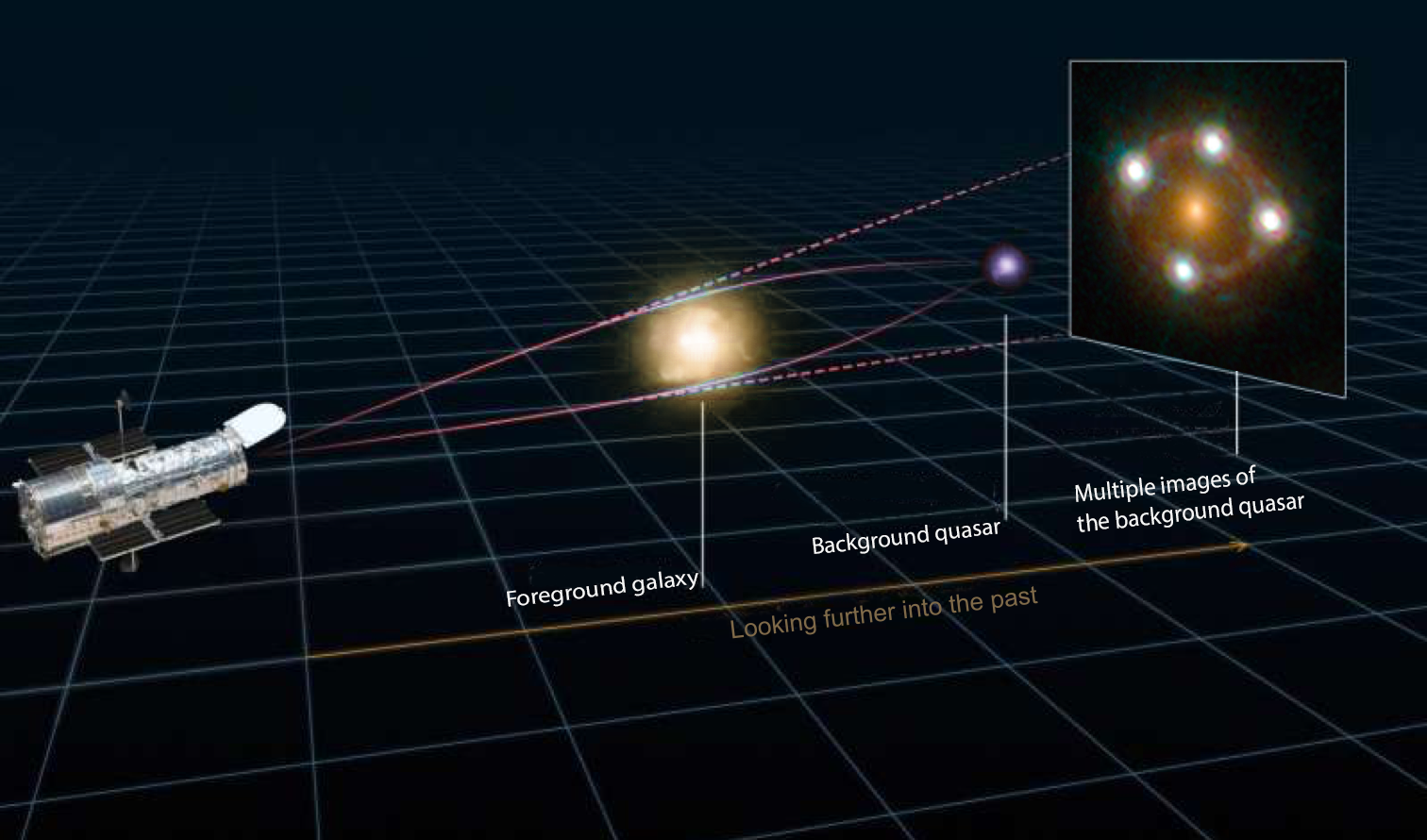}
    \caption{Illustration of the light path of the quadruply imaged lensed quasar HE0435-1223. The different light paths result in different arrival times. The relative time delays between the images is directly proportional to the overall physical distances from the observer to the lens and source. Measuring the time delays and reconstructing the lensing effect allow one to measure an absolute scale in the universe. Graphics from: Martin Millon, Image from Hubble Space Telescope \citep{Wong:2017}.}
    \label{fig:td_overview}
\end{figure*}

When the luminosity of a strongly lensed background source varies over time, such as an active galactic nucleus (AGN), the variability pattern manifests in each of the multiple images and is delayed in time due to the different light paths of the different images (see Figure \ref{fig:td_overview}).
The arrival-time difference between two images $\boldsymbol{\theta}_{\rm A}$ and $\boldsymbol{\theta}_{\rm B}$ that originated from the same source $\boldsymbol{\beta}$, $\Delta t_{\rm AB}$, is
\begin{equation}\label{eqn:time_delay}
    \Delta t_{\rm AB} = \frac{D_{\Delta t}}{c} \left[\tau(\boldsymbol{\theta}_{\rm A}, \boldsymbol{\beta}) - \tau(\boldsymbol{\theta}_{\rm B}, \boldsymbol{\beta}) \right],
\end{equation}
where
\begin{equation}\label{eqn:fermat_potential}
    \tau(\boldsymbol{\theta}, \boldsymbol{\beta}) \equiv \left[ \frac{\left(\boldsymbol{\theta} - \boldsymbol{\beta} \right)^2}{2} - \phi(\boldsymbol{\theta})\right]
\end{equation}
is the Fermat potential \citep{Schneider:1985, Blandford:1986}, and
\begin{equation} \label{eqn:ddt_definition}
    D_{\Delta t} \equiv \left(1 + z_{\rm d}\right) \frac{D_{\rm d}D_{\rm s}}{D_{\rm ds}}
\end{equation}
is the time-delay distance \citep{Refsdal:1964, Schneider:1992, Suyu:2010}. The Fermat potential (Eqn.~\ref{eqn:fermat_potential}) consists of two terms, a geometric term reflecting the geometric path difference, and a potential term, capturing the difference in the local spacetime dilation, known as the Shapiro delay.
The optical terms stated (such as the Fermat potential) are only valid under small-angle and thin-lens assumptions. We refer to \chapintro~ for these assumptions and the more general multi-plane formalism.

\subsection{Angular diameter distances and cosmology}
The angular diameter distance between two redshifts $z_1$ and $z_2$ in an Friedmann--Lemaitre--Robertson--Walker (FLRW) metric is
\begin{equation}
\label{eq:DA}
D(z_1,z_2) = \frac{1}{1+z_2}f_{K}[\chi(z_1,z_2)]
\end{equation}
where
\begin{equation}\label{eqn:chi}
    \chi(z_1,z_2)=\frac{c}{H_0} \int_{z_1}^{z_2} \frac{dz'}{E(z')}
\end{equation}
is the comoving distance with $E(z) \equiv H(z)/H_0$ as the dimensionless Friedman equation %(\textit{ref. section 1.2 `cosmological standard model'})  
and

\begin{equation}
\label{eq:fk}
f_{K}(\chi) = \left\{ \begin{array}{ll}
 K^{-1/2} \sin\left(K^{1/2}\chi\right) & \textrm{for $K>0$}\\
 \chi & \textrm{for $K=0$} \\
 (-K)^{-1/2}\sinh\left[(-K)^{1/2}\chi \right] & \textrm{for $K<0$}
  \end{array} \right.
\end{equation}
is the spatial curvature of the background metric.

In the $\Lambda$CDM cosmology with density parameters $\Omega_{\rm m}$ for matter, $\Omega_{k}$ for spatial curvature, and $\Omega_{\Lambda}$ for dark energy described by the cosmological constant $\Lambda$, the dimensionless Friedman equation ($E(z)$, Eqn.~\ref{eqn:chi}) is given by
\begin{equation}\label{eqn:e_z}
    E(z) = \left(\Omega_{\rm m}(1+z)^3 + \Omega_{k} (1+z)^2 + \Omega_{\Lambda} \right)^{1/2}
\end{equation}
and the spatial curvature is $K=-\Omega_{k}H_0^2/c^2$.

Constraints on the Fermat potential difference $\Delta \tau_{\rm AB}$ and a measured relative time delay $\Delta t_{\rm AB}$ between to images of the same source can be turned into constraints of the time-delay distance $D_{\Delta t}$. $D_{\Delta t}$ is an absolute quantity that has units of distance and anchors the scale of the Universe within the lensing configuration.
The Hubble constant, the local expansion rate of the cosmological background metric, sets the locally linear relationship between relative recession velocities and physical separation of two objects. In the frame of an observer, such as on Earth, for a fixed relative velocity or redshift, the Hubble constant is inversely proportional to the absolute physical distance to the object. 
The Hubble constant is inversely proportional to the absolute scales of objects in the Universe for which redshifts are measured (see e.g., Eqn.~\ref{eqn:chi}) and thus scales with $D_{\Delta t}$ as
\begin{equation} \label{eqn:H0_ddt}
	H_0 \propto D_{\Delta t}^{-1}.
\end{equation}
The direct (inverse) proportionality of the time-delay cosmography measurable quantity $D_{\Delta t}$ and $H_0$ makes $H_0$ the primary cosmological parameter time-delay cosmography can constrain. \footnote{We note that the time-delay distance $D_{\Delta t}$ is not a measurement of $H_0$ at a specific redshift ($z_{\rm lens}$ or $z_{\rm source}$).}
Time-delay cosmography provides primarily an \textit{absolute} distance anchor and hence can provide valuable information to shed light on the current tension in cosmology.
.

There is a secondary mild dependence of the measured Hubble constant when inferred from time-delay cosmography, namely on the relative expansion history from current time ($z=0$) to the redshift of the deflector and the source (dependence on $E(z)$ in Eqn.~\ref{eqn:e_z}).
The mild dependency on other cosmological parameters beyond $H_0$ can be compensated with other cosmological probes that are sensitive to the \textit{relative} expansion history (such as SNIa luminosity distances, e.g., \citealt{Taubenberger19, Arendse19, Liao:2019, Liao:2020}), or with a large set of gravitational lenses at different lens and source redshifts \citep[e.g.,][]{Li2024}.

\subsection{Observables and degeneracies}

The time delay between two images $\Delta t_{\rm AB}$ can be measured from light curves and is hence a direct observable (see Section~\ref{sec7:time_delays}). The relative Fermat potential $\Delta \tau_{\rm AB}$, however, is not a direct observable. The primary observations used to infer $\Delta \tau_{\rm AB}$ are positional constraints of multiple imaged sources and their extended distortions in the lensed arcs from the lensing effect. However, there are degeneracies inherent in gravitational lensing that limit the amount of information accessible by positional and distortion effects as observed in imaging data \citep[e.g.,][]{Falco:1985, Gorenstein:1988, Kochanek:2002, Saha:2006, Schneider:2013, Schneider:2014, Birrer:2016, Unruh:2017, Birrer:2021curvedarcs} and \chapintro.

The mass-sheet degeneracy \citep[MSD;][]{Falco:1985} is the most prominent lensing degeneracy impacting the prediction of the Fermat potential and hence time-delay cosmography.
The MSD stems formally from a multiplicative transform of the lens equation (\eqref{eqn:lens_equation}) which preserves image positions under a linear source displacement $\boldsymbol{\beta} \rightarrow \lambda\boldsymbol{\beta}$ 
combined with a transformation of the convergence field
\begin{equation}\label{eqn:mst}
    \kappa_{\lambda}(\boldsymbol{\theta}) = \lambda \kappa(\boldsymbol{\theta}) + \left( 1 - \lambda\right).
\end{equation}
\eqref{eqn:mst} above is known as the mass sheet transform (MST) and is a \textit{mathematical transformation} where the term $(1 - \lambda) \equiv \kappa_{\rm MST}$ is equivalent to a uniform-surface density sheet of convergence (or mass) that extends to infinite angular scales, and hence the name mass sheet transform and the name of the experienced degeneracy, the mass sheet degeneracy. $\kappa_{\rm MST}$ can be positive or negative, since it is defined relative relative to the average positive density of the universe.
The MST, by means of preserving image positions and being linear, also preserves any higher order \textit{relative} differentials of the lens equation. Absolute lensing quantities, such as absolute magnification or size of structure, however, are not preserved by the MST.
Only observables related to either the unlensed apparent source size ($\boldsymbol{\beta}$ vs. $\lambda\boldsymbol{\beta}$), such as the unlensed apparent brightness, or the lensing potential are able to break the MSD.
For example, the same relative lensing observables can be predicted if the mass profile is scaled by the factor $\lambda$ with the addition of a sheet of convergence (or mass) of $\kappa_{\rm MST}(\boldsymbol{\theta}) = (1-\lambda)$ and re-sizing of the source by a factor $\lambda$.

The Fermat potential difference between a pair of images A\& B (\eqref{eqn:fermat_potential}) scales with $\lambda$ as
\begin{equation}\label{eqn:fermat_mst}
    \Delta \tau_{\rm AB , \lambda} =  \lambda \Delta \tau_{\rm AB},
\end{equation}
and so does the relative time delay as
\begin{equation}\label{eqn:time_delay_mst}
    \Delta t_{\rm AB , \lambda} =  \lambda \Delta t_{\rm AB}.
\end{equation}
When transforming a lens model with an MST, the inference of the time-delay distance (Eqn.~\ref{eqn:ddt_definition}) from a measured time delay and previously inferred Fermat potential transforms as
\begin{equation} \label{eqn:ddt_mst}
    D_{\Delta t , \lambda} = \lambda^{-1}D_{\Delta t}.
\end{equation}
In turn, the Hubble constant, when inferred from the time-delay distance $D_{\Delta t}$, transforms as (from Eqn.~\ref{eqn:H0_ddt})
\begin{equation} \label{eqn:h0_mst}
H_{0 , \lambda} =  \lambda H_0.
\end{equation}
% Point out that time-delay distance is not a measurement of H0 at a specific redshift (z_lens or z_source).  This is a common mistake that I see in the literature.

An MSD effect relative to a specified deflector model might be associated with the mass distribution of the main deflector, referred as \textit{internal} MSD with $\lambda_{\rm int}$, or with inhomogenities along the line of sight (LOS) of the strong lens system, referred as \textit{external} MSD.

Mass over- or under-densities relative to the mean background density along the LOS of the strong lensing system cause, to first order, shear and convergence lensing perturbations \citep[e.g.,][]{Dalal:2005, Hilbert:2007, PuchweinHilbert:2009}. Reduced shear distortions do have a measurable imprint on the azimuthal structure of the strong lensing arcs \citep[see e.g.,][]{Birrer:2021curvedarcs, Hogg:2022}. In contrast, the convergence component of the LOS, denoted as $\kappa_{\rm ext}$, describes the focusing or de-focusing of the light rays and is equivalent to an MST, $\kappa_{\rm ext} \equiv (1 - \lambda)$, and hence not directly measurable from imaging data.

Equivalent to describing the (de-) focusing along specific line-of-sights by convergence terms, we can alter the specific angular diameter distance relative to the homogeneous background metric.
In our notation, $D^{\rm lens}$ is the angular diameter distance along a specific line of sight, including all structure such as LOS. $D^{\rm bkg}$ is the angular diameter distance just from the homogeneous background metric without any perturbative contribution. The relation between $D^{\rm lens}$ and $D^{\rm bkg}$ are given by the convergence terms as
\begin{multline}\label{eqn:ang_distance_kappa}
D^{\rm lens}_{\rm d} = (1 - \kappa_{\rm d})D_{\rm d}^{\rm bkg}\\
D^{\rm lens}_{\rm s} = (1 - \kappa_{\rm s})D_{\rm s}^{\rm bkg}\\
D^{\rm lens}_{\rm ds} = (1 - \kappa_{\rm ds})D_{\rm ds}^{\rm bkg},\\
\end{multline}
where $\kappa_{\rm d}$ is the external convergence from the observer to the deflector, $\kappa_{\rm s}$ from the observer to the source, and $\kappa_{\rm ds}$ from the deflector to the source, respectively \citep{Birrer:tdcosmoiv}.
The individual convergence terms can be calculated in the Born approximation along undeflected light paths independent of the strong lensing deflector.

The notation of perturbed angular diameter distances allow us also to directly calculate the impact of line-of-sight structure on the cosmographic inference, and in particular the measurement of the Hubble constant.
The time delay can be described as the product of three different angular diameter distances entering $D_{\Delta t}$ in Equation~\ref{eqn:ddt_definition} \citep{Birrer:tdcosmoiv, Fleury:2021metric}, and hence the effective external convergence $\kappa_{\rm ext}$impacting the time delay and time-delay distance is
\begin{equation}\label{eqn:mst_combined}
    1 - \kappa_{\rm ext} = \frac{(1 - \kappa_{\rm d})(1 - \kappa_{\rm s})}{1 - \kappa_{\rm ds}}.
\end{equation}
% metric distortions of LOS structure
We note that, also directly visible from the equation above,  
the lensing efficiency (see~\chapintro) impacting the linear distortions for both shear and $\kappa_{\rm ext}$ is different from the standard weak lensing efficiency in the absence of a strong lensing deflector \citep{McCully:2014, McCully:2017, Birrer:2017los, Birrer:tdcosmoiv, Fleury:2021los}. 

% internal MST
Uncertainties or biases related to the MSD may also arise in regards to assumptions made in the radial density profile of the main deflector galaxy \citep[see e.g.][]{Kochanek:2002, Saha:2006, Read:2007, Schneider:2013, Coles:2014, Xu:2016, Birrer:2016, Unruh:2017, Sonnenfeld:2018, Kochanek:2020, Blum:2020, Birrer:tdcosmoiv, Kochanek:2021}. 

We refer to the MSD attributed to uncertainties in the radial density profile as the \textit{internal} MSD, and describe its effect with the MST parameter
$\lambda_{\rm int}$ relative to an assumed mass profile. We will further discuss this aspect in Section~\ref{sec7:mass_profile}.

The total MST, the relevant transform to constrain for an accurate Fermat potential determination and $H_0$ measurement, is the product of the internal and external MST \citep[e.g.,][]{Schneider:2013, Birrer:2016, Birrer:tdcosmoiv}
\begin{equation}\label{eqn:lambda_combined}
    \lambda = (1-\kappa_{\rm ext}) \times \lambda_{\rm int}.
\end{equation}

To summarize, the prediction of the time delay (Equation~\ref{eqn:time_delay} can be generalized to \begin{equation}\label{eqn:time_delay_generalized}
    \Delta t_{\rm AB} = (1-\kappa_{\rm ext}) \lambda_{\rm int} \frac{\tdist}{c} \Delta \tau_{\rm AB}.
\end{equation}

The existence of the MST and its generalizations imply that one has to rely either on (1) non-lensing information that can specifically constrain the shape of the radial mass density profile or (2) assumptions and priors about the functional form of the radial mass density profile that limit the degrees of freedom in the direction of the MST to constrain the lens model with a sufficient level of precision for time-delay cosmography.

The line-of-sight lensing contribution, $\kappa_{\rm ext}$, can be estimated by tracers of the large-scale structure using galaxy number counts \citep[e.g.,][]{Greene:2013, Rusu:2017} or weak-lensing measurements \citep{Tihhonova:2018, Tihhonova:2020}. Galaxy number counts, paired with a cosmological model including a galaxy--halo connection, are able to constrain the probability distribution of $\kappa_{\rm ext}$ to a few per cent per sight line. The main uncertainty in this approach arise from the uncertainties of luminous matter tracing the more dominant dark matter structure.

To break the total MSD $\lambda$ with observations, we require observations that are sensitive to the total MSD. Stellar kinematics is the most prominent and commonly used one to break the total MSD. The collective motion of stars is a direct tracer of the three-dimensional gravitational potential and hence provides an independent mass estimate.
Joint lensing+dynamics constraints have been used to provide measurements of galaxy mass profiles \citep[e.g.,][]{Grogin:1996, Romanowsky:1999, Treu:2002, Koopmans:2004, Barnabe:2011, Barnabe:2012}.
The modelling of the kinematic observables in lensing galaxies range in complexity from spherical Jeans modeling \citep{BinneyTremaine:2008} to Schwarzschild  \citep{Schwarzschild:1979} methods.

The prediction of the LOS velocity dispersion $\sigma_{\rm v}$ from any model, regardless of the approach, can be decomposed into a cosmology-dependent and cosmology-independent part, as \citep[see e.g.,][]{Birrer:2016, Birrer:2019}
\begin{equation}\label{eqn:los_sigma_v}
    \sigma_{\rm v}^2 =  \frac{1-\kappa_{\rm s}}{1 - \kappa_{\rm ds}} \frac{D_{\rm s}}{D_{\rm ds}}c^2 J(\boldsymbol{\xi}_{\rm lens}, \boldsymbol{\beta}_{\rm ani},\lambda_{\rm int}),
\end{equation}
where $J$ is a dimensionless quantity dependent on the deflector model parameters ($\boldsymbol{\xi}_{\rm lens}$),  $c$ is the speed of light, and $\boldsymbol{\beta}_{\rm ani}$ the stellar anisotropy distribution. The dimensionless factor $J$ incorporates also  the observational conditions and luminosity-weighting within the aperture of the dispersion measurement being taken \citep[e.g.,][]{Binney:1982, Treu:2004, Suyu:2010}.
The internal component $\lambda_{\rm int}$ should be physically interpretable as a three-dimensional mass profile and incorporated into the kinematics modeling term $J$, in particular when there are multiple aperture measurements available \citep{Teodori:2022}.
In the approximate case of a very extended sheet-like perturbation, we can approximate
\begin{equation}\label{eqn:j_lambda_int}
    J(\boldsymbol{\xi}_{\rm lens}, \boldsymbol{\beta}_{\rm ani},\lambda_{\rm int}) \approx \lambda_{\rm int} J(\boldsymbol{\xi}_{\rm lens}, \boldsymbol{\beta}_{\rm ani}).
\end{equation}

Combined lensing+dynamics constraints are sensitive to the combination of terms present in Equation~\ref{eqn:los_sigma_v}. Only the combination of terms, i.e., $\lambda_{\rm int})$, $D_{\rm s}$, $D_{\rm ds}$, $\boldsymbol{\beta}_{\rm ani}$, $\kappa_{\rm s}$, $\kappa_{\rm ds}$, is constrained and hence assumptions or priors on parts of the terms are required to provide a precise statement other terms.
For example, when assuming the relative expansion history through the involved angular diameter distance ratio $D_{\rm s}/D_{\rm ds}$, and the LOS contributions $\kappa_{\rm s}$ and $\kappa_{\rm ds}$, an inference on $\lambda_{\rm int}$ is possible. On the other hand, when assuming $\lambda_{\rm int}$ and the convergence terms, an inference on the angular diameter distance ratio, and hence the relative expansion history, is possible.

When combining time delays with lensing+dynamics, the observations of the time delay and kinematics need to be simultaneously be described by Equation~\ref{eqn:time_delay_generalized} and Equation~\ref{eqn:los_sigma_v} in addition to the imaging data. These two independent equations can be arbitrarily algebraically combined in two-dimensional angular diameter distance constraints \citep{Birrer:2016, Birrer:2019}. A convenient transform of those constraints is in the basis of
\begin{equation}\label{eqn:Ddt_cosmography}
    \tdist = \frac{1}{(1-\kappa_{\rm ext}) \lambda_{\rm int}}\frac{c \Delta t_{\rm AB}}{\Delta \tau_{\rm AB}}
\end{equation}
and
\begin{equation}\label{eqn:Dd_cosmography}
    D_{\rm d} = \frac{1}{1 - z_{\rm d}}\frac{1}{1 - \kappa_{\rm d}}\frac{c \Delta t_{\rm AB}}{\Delta \tau_{\rm AB}}  \frac{c^2 J(\boldsymbol{\xi}_{\rm lens}, \boldsymbol{\beta}_{\rm ani}, \lambda_{\rm int})}{\lambda_{\rm int}\sigma^2_{\rm v}}.
\end{equation}

When mapped into the $D_{\Delta t}$--$D_{\rm d}$ plane as outlined above, the projection on constraints in $D_{\rm d}$ is invariant under any pure external MSD parameter $\kext$
\citep{Paraficz:2009, Jee:2015, Birrer:2019, Yildirim:2021}\footnote{$D_{\rm d}$ is still dependent on the LOS between observer and lens, $\kappa_{\rm d}$ (Eqn. \ref{eqn:ang_distance_kappa}).}.
If the approximation of Equation~\ref{eqn:j_lambda_int} holds, $D_{\rm d}$ becomes even independent of $\lambda_{\rm int}$, and is overall less susceptible to the internal MSD.

%%%%%%%%

\section{Overview of analysis ingredients}\label{sec7:analysis_overview}

To measure the cosmographic distances, in particular the time-delay distance $D_{\Delta t}$ (Equation~\ref{eqn:ddt_definition}), or the more general $D_{\Delta t}$--$D_{\rm d}$ (Equations~\ref{eqn:Ddt_cosmography}-\ref{eqn:Dd_cosmography}) combination, from a strong lensing system with a time-variable source, the following data products are required:
\begin{enumerate}
\item discovery of a lens with a time-variable source that is multiply imaged,
\item spectroscopic redshifts of the source, $z_{\rm s}$, and lens, $z_{\rm d}$,
\item measured time delays between at least one multiple image pair,
\item lens mass model to determine the Fermat potential between the multiple images from sufficiently high resolution imaging to resolve the positions of the multiply-imaged quasars,
\item lens environment studies to constrain external lensing effects.
\end{enumerate}

A complete analysis for an individual lensing system requires the coordination of multiple independent observations. 
The analysis can be severely limited in its precision and reliability due to a single missing ingredient.
For example, without measurements of a time delay, no constraints on absolute distances involved can be achieved, and thus, regardless of the approach or external priors chosen, no direct constraints on the Hubble constant can be made.

The spectroscopic redshifts of the quasar sources, $\zs$, are often obtained using the frequent emission lines in quasars. The redshift of the lens, $\zd$, can be challenging to measure since since massive elliptical galaxies lack prominent and sharp absorption or emission lines and the bright quasar images can outshine the lens galaxy. Measuring $\zd$ of a lensed quasar systems often require high signal-to-noise ratio spectra taken under good seeing conditions, to deblend the lensing galaxy from the quasar. Technically, the redshifts involved in the lensing system are not directly required for the distance measurement. However, for the cosmological interpretation of the obtained distances, the redshifts are of crucial importance.

We describe in the next sections the remaining three ingredients; time delays (Section~\ref{sec7:time_delays}), lensing potential (galaxy scale and cluster) (Section~\ref{sec7:lens_potential}), and line-of-sight perturbations (Section~\ref{sec7:los}).

%%%%%%%%

\section{Measuring time delays}\label{sec7:time_delays}

\subsection{Monitoring of lensed quasars}

Lensed quasars are variable on short timescale, making the time-delay measurements possible, and  sufficiently bright to be observed at cosmological distances. They were hence quickly identified as excellent sources for time-delay cosmography. Lensed quasars are also currently much more common than lensed supernovae as around 300 lensed quasars have been discovered at the time of writing compared to only four lensed supernovae (see \chapsn). Lensed quasars are typically found in the redshift range $\zs~\sim~1-3$, with massive early-type galaxies acting as the lenses located around redshift $\zd \sim 0.2-0.8$ \citep[e.g.,][and \chapsearch]{Lemon:2022}. This lensing configuration typically produces multiple images separated by a few arcseconds, which is sufficient to be resolved with small ground-based telescope, Typical time delays in this configuration are of the order of days to a year. The monitoring of lensed quasars and the measurement of their individual brightness fluctuations is thus challenging but possible with 1-m or 2-m class telescopes, provided that a regular and long-term access is guaranteed \citep[see e.g. the COSMOGRAIL collaboration][]{Eigenbrod2005, Courbin2011, Tewes2013a}. The relative error on the time delays, which is directly propagated to $H_0$, depends on the absolute errors divided by the time-delay itself. Therefore, long-delay lenses are more valuable for cosmography as they yield smaller $H_0$ uncertainties from this component. The achievable precision on time-delay measurements is limited by several astrophysical, observational and instrumental factors, that are listed below.

\paragraph{Photometric accuracy :} In the optical, most quasars are variable on a timescale of weeks to years, and the longest variations also have the largest amplitude. This means that either long-duration light curves or high photometric accuracy are required to measure the delay reliably. In one visibility season, variations of the order of 0.2\,mag are often observed, which requires a photometric accuracy of a few milli-magnitudes to identify precisely the inflection points. These inflection points are essential features in the light curves since it is not possible to measure a time delay if the quasars does not display any variations, or if the first derivative remains always constant. Reaching a photometric accuracy of only a few milli-magnitude is challenging as the quasar images are often blended with extended sources such as gravitational arcs or the lens galaxy. Consequently, the reconstruction of the Point Spread Function (PSF) and proper treatment of the contaminating light from these extended sources are usually the key to reduce the noise in the light curves. 

\begin{figure*}[htbp]
\centering
    \includegraphics[width=0.8\textwidth]{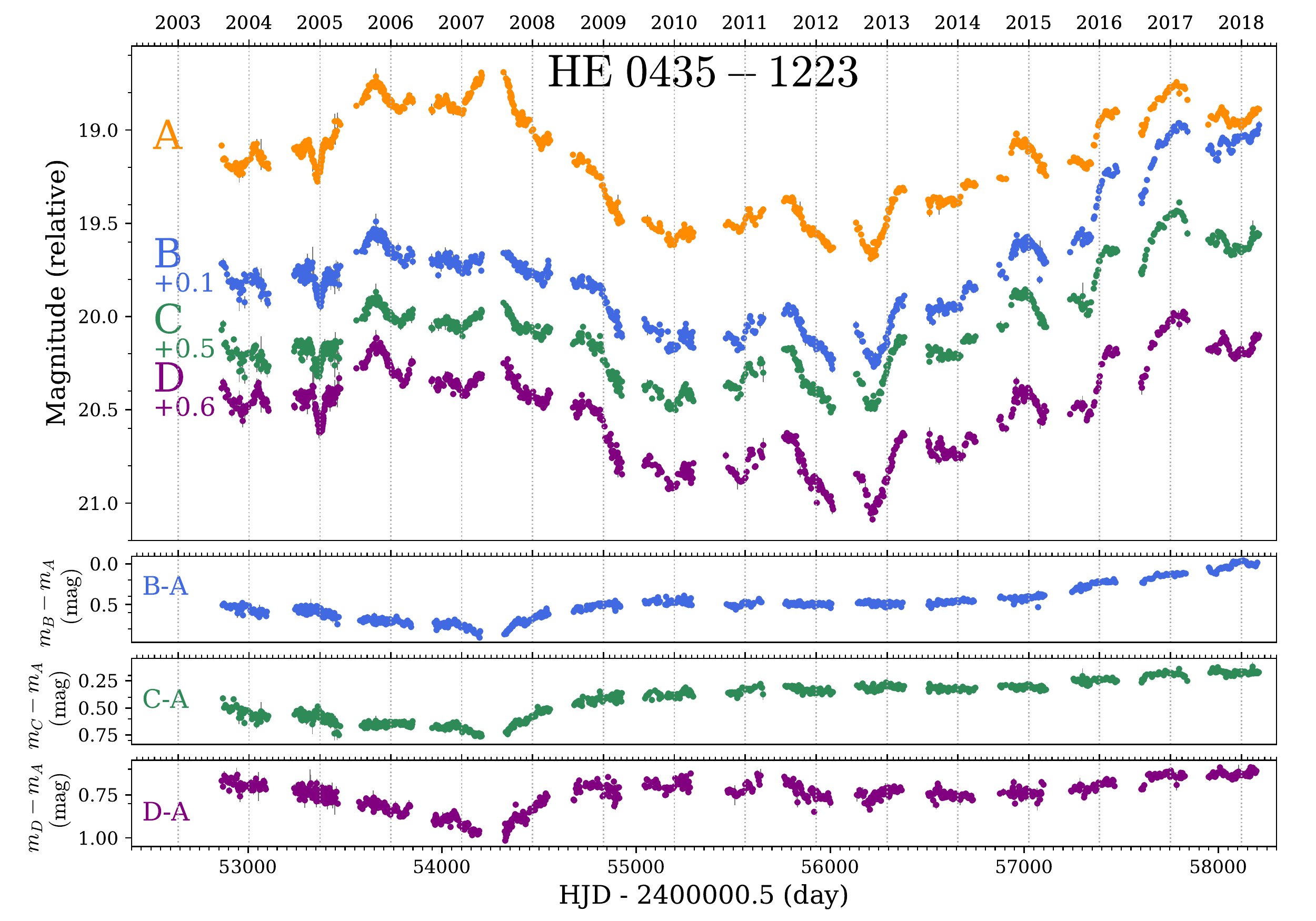}
    \caption{R-band light curves of the lensed quasar HE0435-1223, obtained by the COSMOGRAIL program from the Euler 1.2m Swiss Telescope, the 1.5m telescope at the Maidanak Observatory, the Mercator 1.2m telescope, and the SMARTS 1.3m telescope. The bottom panels corresponds to the difference between pairs of light curves, corrected by the time delays, highlighting the microlensing variability. Figure reproduced from \citep{Millon2020b}. }
    \label{fig:H0435_lcs}
\end{figure*}

\paragraph{Monitoring cadence and duration of the monitoring:} A fast and precise temporal sampling of the light curve is necessary if one targets the fast variations of small amplitudes of the quasar.
The monitoring cadence then needs to be commensurate with the timescale of the targeted variations.
The total time span of the monitoring campaign also needs to be sufficient to cover the lensing time delays and to ensure that enough variations of the quasar are recorded for multiple images at relative delayed times. To obtain light curves to such specifications requires continuous access to the telescope for at least one visibility season, which is typically 6 to 8 months.

\paragraph{Windowing effects and correlated noise:} Seasonal gaps are often unavoidable in optical light curves since only circumpolar targets are observable all year-long. The fact that data are missing every year can introduce some windowing effects, which should be accounted for when using cross-cor\-relation techniques to measure time delays. The missing data introduces a periodic signal that must be carefully removed or taken into account before attempting to measure the time delays. Additionally, great care should be taken in the presence of correlated noise, which is often present in the light curves due to uncertainties in the assignment of flux coming from different quasar images. If no evident variations can be matched unambiguously in both light curves, it is unlikely that any statistical methods will robustly measure a time delay.

\paragraph{Extrinsic variability:} Extrinsic variations are often observed in the light curves. They are caused mainly by the microlensing of the quasar images, and also a variety of other astrophysical effects \citep [see e.g.][and \chapmicro]{Schechter2003, Blackburne2010, Dexter2011, bib7:Sluse2014}. Microlensing is caused by the stars in the lensing galaxy, which add some extra time-variable "micro-magnification" on top of the static "macro-magnification" produced by the lensing galaxy. As described in \chapmicro, the modulation of the micro-magni\-fica\-tion due to the relative motion between the quasar, the lens and the observer introduces some extrinsic variations on top of the quasar intrinsic variations. For this reason, the light curves, even shifted in time and magnitude, rarely match perfectly. These extrinsic variations can severely bias time-delay measurements if not appropriately modelled and marginalized over. \\

In the past two decades, these difficulties have been progressively dealt with. The advances in photometric instrumentation in the late 1990s allowed us to acquire accurate and well-sampled light curves, which yielded the first robust time-delay measurements from optical monitoring \citep[e.g.,][]{Kundic1997,  Schechter1997, Burud2000, Burud2002, Hjorth2002, Colley2003, Kochanek2006} and in radio monitoring \citep{Fassnacht1999, Fassnacht2002, Biggs1999, Koopmans:2000}. Although some of these measurements already reached an excellent precision of a few percents, the majority had $\sim$10-15\% errors, hence limiting the measurement of $H_0$ to the same precision. 
These first encouraging results led to a systematic attempt to monitor a sample of lensed quasars in both hemispheres by the COSMOGRAIL program, which started in 2003 \citep{Courbin2005}. The observing strategy then was to follow a dozen of lenses at bi-weekly cadence until the time delays can be measured to a few percent precision. This strategy yielded precise measurements for the brightest and most variable objects in about five years \citep{Vuissoz2007,Vuissoz2008,Courbin2011,Tewes2013b,Eulaers2013,Rathna2013} but required more than a decade of monitoring to obtain the time delays for most of the less variable and fainter targets \citep{Millon2020b}. An example of a light curves acquired by the COSMOGRAIL program over the past 15 years is shown in Figure \ref{fig:H0435_lcs}. Thanks to this long-term observing effort and other monitoring campaigns \citep[e.g.][]{Poindexter2007, Goicoechea2016, Giannini2017, Shalyapin2019}, about 40 lensed quasars have now known time delays, although with variable precision, but the sample starts to be sufficiently large to vastly reduce the random uncertainties and to enable a statistical study of the time-delay lenses. In addition, three cluster-scale lensed quasars have measured time delays \citep{Fohlmeister2007, Fohlmeister2008, Fohlmeister2013, Dahle2015, Munoz2022} but their modelling is much more complex than galaxy-scale lenses (see Section \ref{sec:cluster_h0} for detail).

\begin{figure*}[htbp]
\centering
    \includegraphics[width=0.9\textwidth]{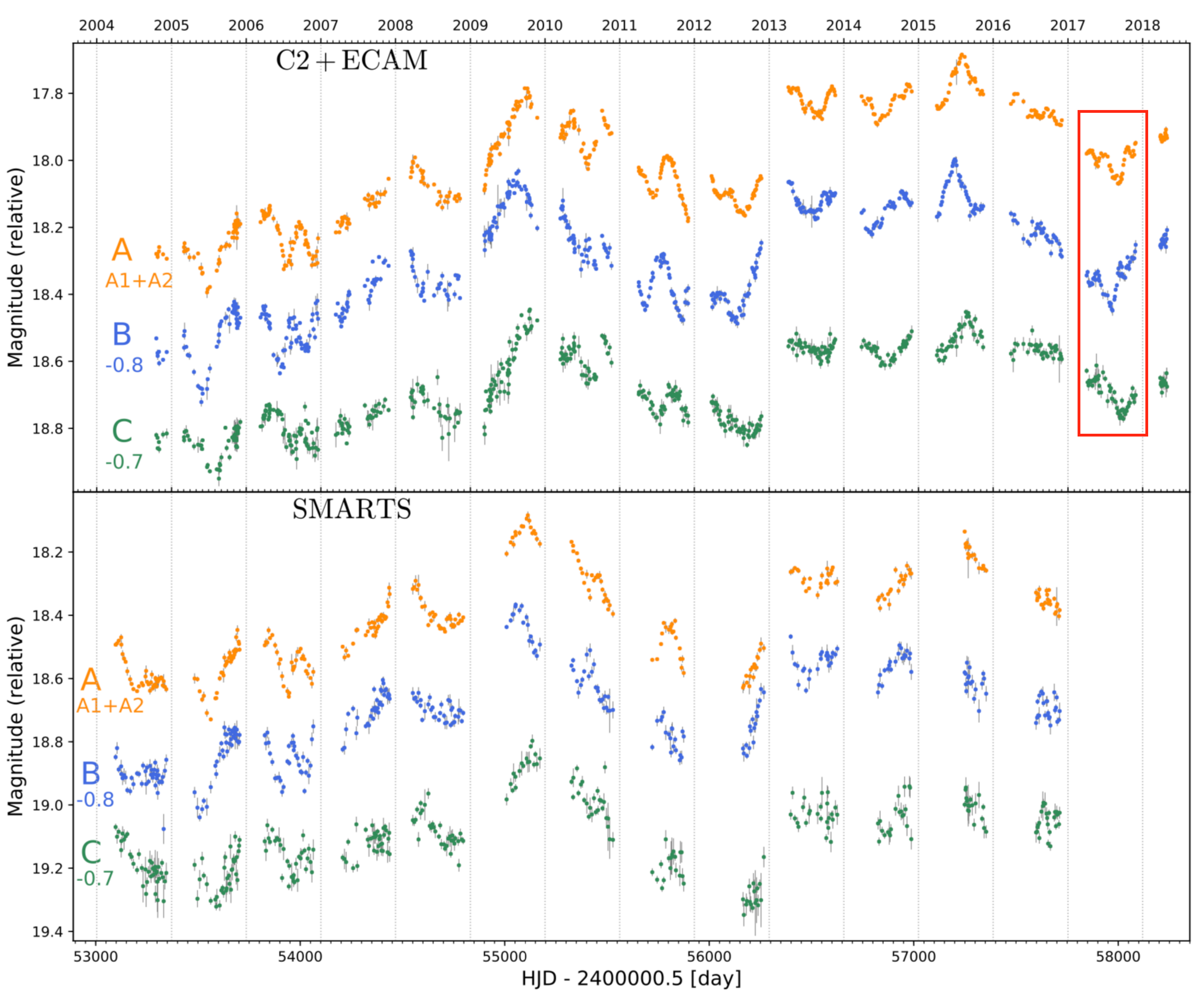}
    \vskip 10pt
    \includegraphics[width=0.5\textwidth]{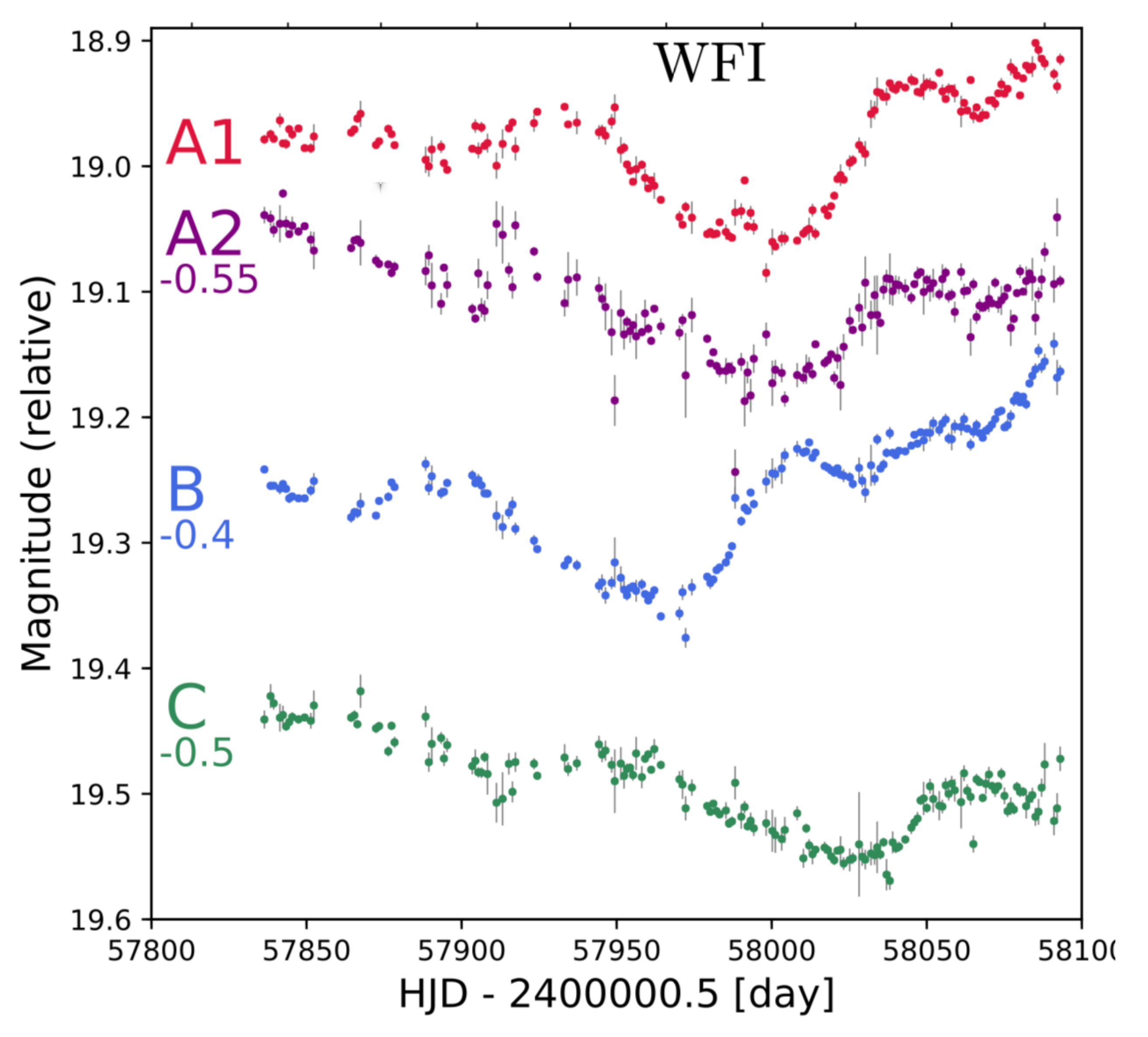}
    \caption{Comparison of two observing cadences. On the two top panels are shown 14-year-long light curves of WFI~2033-4723, observed with a 4-day cadence at the 1.2m Euler telescope at La Silla and at the SMARTS telescope at Las Campanas. During the season indicated with a red rectangle, the object has also been observed daily with the MPIA 2.2m at La Silla (bottom panel), unveiling exquisite small-scale structures that vary faster than microlensing. Such observations allow to measure time delays in 1 single season, with similar accuracy and precision than the lower cadence data over 14 years \citep[]{Bonvin2019, Millon2020b}.}
    \label{fig:2033}
\end{figure*}

As time-delay cosmography is now entering a new regime with an increasing number of lensed quasars being discovered every year, the time delays now need to be measured rapidly to turn these newly discovered systems into cosmological constraints. \cite{Courbin2018} demonstrated that it is possible to obtain accurate time-delay measurements in only one monitoring season thanks to the small amplitude variations of the quasars, of the order of 10 to 50 millimag, that happen on a timescale of weeks or months. These variations are faster than the microlensing variability, which varies on a typical timescale of several months or years. If detected at a sufficient signal-to-noise ratio (SNR), these features reduce the need for long light curves as it allows us to disentangle the intrinsic and microlensing variability more easily. However, this change of strategy requires almost a daily cadence to obtain a sufficient sampling of these small features in the light curves. Their amplitude is of the order of 10 mmag, which requires 2-m class telescopes to obtain a sufficient SNR in 30 minutes of exposure at magnitude as faint as $\sim 20$. This is illustrated in Fig.~\ref{fig:2033} in the case of the bright quadruple quasar WFI~2033-4723 \citep[]{Bonvin2019}. The technique enabled to measure six new time delays in one single season \citep{millon2020c}, with more to follow. 

In the future, the Vera Rubin Observatory will obtain high-SNR multiband data for all southern lensed quasars, opening the possibility of building a sample of a few hundreds lensed quasars with known time delays. The cadence will however be limited to one point every few days in each band, which might not be sufficient to obtain the time delays at a few percent precision for the most interesting targets. Complementary observations from a 2-m class telescope at a daily cadence might still be useful to refine the time-delay measurements of the most promising objects.

\subsection{Time-delay measurements techniques}

Once well-sampled light-curves have been acquired, the next step consists of identifying time-variable features that can be matched in all light curves from the individual images. The shifting of the light curve to match the features leads to a measurement of the time delays.

This step is significantly complicated by the presence of extrinsic variations due to microlensing in the light curves on top of the quasar intrinsic variations. The signature of micro-lensing can be seen in most lensed quasar light curves. In some cases, it manifests itself by a sharp rise of the luminosity in one of the multiple images, which happens when the source approaches or crosses a micro-caustic. This probably happened, for example, in 2007 for image A of HE0435$-$1223. In Figure \ref{fig:H0435_lcs}, we show the light curves of HE0435$-$1223 as well as the differential curves between the images, highlighting the microlensing signal. Caustic crossing events are not the only signature of microlensing visible in the data. Most quasar light curves also exhibit slow variations of the microlensing over several years. An example of this phenomenon is image B of HE0435$-$1223, which slowly raised by $\sim$~0.5 mag between 2013 and 2018. It typically happens when the stellar density is high and the quasar is located in regions where the microcaustics overlap. The net effect is a smooth variation of the microlensing magnification as the quasar moves through these crowded regions. The extrinsic variations introduced by microlensing contains valuable information about the quasar accretion disk structure but is a real source of nuisance when measuring the time delays from optical monitoring. However, radio light curves are generally less influenced by microlensing compared to their optical counterparts. This is because the region from which radio waves are emitted is usually significantly larger than the microcaustics, so the impact of microlensing tends to be averaged out and less noticeable. In the optical range, the accretion disk, responsible for most of the UV and optical emission, has a comparable size to the microcaustics. Consequently, optical light curves are more affected by microlensing, leading to important extrinsic variations (for a detailed description of how microlensing differently impacts various emission regions, refer to \chapmicro).

To deal with this issue, several curve-shifting algorithms have been proposed over the years, which can be classified into two categories. On one hand, some methods are based on the light-curve cross-correlations \citep[e.g.][]{Pelt1996}, sometimes without attempting to subtract the microlensing variability \citep[e.g. the smoothing and cross-correlation method by][]{Aghamousa2015}. On the other hand, several techniques rely on the analytical modelling of the intrinsic variability of the quasars and/or microlensing variations with, for example, splines \citep{Tewes2013a} or Gaussian Processes \citep[e.g.][]{Hojjati2013}. When explicitly modelled, the microlensing variations are removed from the light curves before attempting to find the optimal time delays. Due to the broad band nature of the monitored signal, mixing flux arising from multiple emission regions, microlensing is rarely perfectly removed but this is shown to have in general a negligible impact on the delay \citep{bib7:Sluse2014}. One can also mention the recent work by \cite{Taak2016}, \cite{Donnan2021} and \cite{Meyer2022}, aiming to infer the time delays in a Bayesian framework, including an explicit modelling of the microlensing variations. 

These methods were tested in the "Time Delay Challenge"\cite[TDC;][]{Dobler2015}, a blind data challenge aiming at assessing the precision and accuracy of the curve shifting algorithm on simulated but realistic data, which includes the microlensing variability. The results and conclusions of the challenge are presented in \cite{Liao2015} as well as in individual papers \citep[][]{Hojjati2014, Bonvin2016}. The problem is in fact more complicated than it sounds since a large fraction of the participating teams did not meet the requirements in term of precision and accuracy on the first and simplest rung of the challenge. Among the qualified teams to participate to the more advanced rungs of the TDC, the different proposed techniques showed overall good performance given the actual quality of the data. Several teams reached an accuracy of $\lesssim $1\% on the most variable light curves. However, it remains to be checked if this level of performance holds if more realistic accretion disk emission mechanisms and source-size effects are included in the simulations. 

Among these source-size effects, microlensing time delay, which is described in details in \cite{TieKochanek:2018}, may be a more subtle manifestation of microlensing acting as a nuisance to measure the time delay. Although it has never been detected so far in lensed quasar light curves directly, this effect may arise when different emission regions of the accretion disc are differentially magnified. A simple model to explain the UV and optical variability of the quasars is the "lamp post" model \citep[e.g.][]{Cackett2007, Starkey2017}, where the luminosity fluctuations originate close from the supermassive black hole and then illuminate the rest of the disk. This triggers temperature fluctuations in the disk, which result into delayed UV and optical emission due to the light travel time from the center. In the absence of differential magnification caused by microlensing, this time lag cancels out between the multiple images and only the "cosmological" time delay is observed. However, if one of the multiple image is affected by microlensing, the time lag originating from a particular region of the disk might be amplified by microlensing and a net excess of microlensing time delay can add to the "cosmological" time-delay. This effect could reach a few hours to a couple of days, which is negligible for most of the systems with long time delays but it can significantly increase the uncertainties for systems with short time delays. This effect can however be mitigated with multi-band light curves \citep{Chan2021, Liao:2020microlensing, Liao:2021} or a proper Bayesian treatment of this effect as a source of nuisance \citep{Chen2018}.  

Future developments of curve shifting algorithms might also include time-delay measurements from unresolved light curves \citep[e.g.][]{Hirv2007, Shu2021, Biggio2021, Springer2021, Bag:2023}, which will open the possibility to monitor small-separation ($<1$\arcsec) lensed quasars. While precise delays from unresolved lightcurves have already been measured in the gamma-ray range \citep{Barnacka:2011, Cheung:2014}, they cannot be used for cosmography as the location of the gamma-ray emission w.r.t. the central AGN remains unknown. In the optical range, space-based large sky surveys, such as Euclid, are expected to discover thousands of small-separation systems, which will not be fully resolved from the ground with our current follow-up facilities. To turn this large sample of small-separation lensed quasars into a useful cosmological probe will require to develop these new techniques. 

%%%%%%%%

\section{Determining lensing potential}\label{sec7:lens_potential}
Determining the lensing potential $\phi$ is a crucial ingredient in time-delay cosmography as it directly enters the time-delay prediction through the Fermat potential (Equation \ref{eqn:fermat_potential}).
The Fermat potential is generally dominated by a massive elliptical galaxy acting as the main deflector and intervening line-of-sight over- and under-densities. To achieve a precise and accurate cosmographic inference, knowledge of both the line-of-sight structure and the mass distribution within the main deflector need to be known.
One of the major limitations in a precise determination of the Fermat potential is the MSD, and the inability with imaging data to constrain the Fermat potential. Thus, either physical assumptions based on what we know from other modelled galaxies, e.g. in the nearby universe, or external data, such as stellar kinematics, is required to constrain the Fermat potential. The measurement of the Hubble constant and constraining the galaxy density profiles are tightly connected and most of the questions asked and techniques being used in \chapgal~ are of the same relevance and applicability for time-delay cosmography.

We discuss observables and inferences, first, for historical context on positional constraints alone in Section~\ref{sec7:conjugate_point}, and then from imaging data in Section~\ref{sec7:imaging_data}.
We then discuss assumptions on mass profiles in Section~\ref{sec7:mass_profile} and what external information provide necessary constraints in Section~\ref{sec7:external_data}.

\subsection{Conjugate point analysis}\label{sec7:conjugate_point}
Historically, the first lens models constructed for time-delay cosmography were based on positional constraints of the quasar images alone, and in combination with the time delays, and to some extent the image magnifications. Valid models must satisfy the lens equation (Eqn.~\ref{eqn:lens_equation}) such that all images must map back to the same source position. In the case of a quadruply lensed system, the four image positions result in constraints of five relative distortion angles (plus translation and rotation of the system). With such limited information, investigators had to assume a very simple form for the lens mass distribution, such as a singular isothermal sphere \citep{Koopmans:1999}, for meaningful constraints on the model.
Different assumptions on the mass profiles lead to vastly different results
\citep[see e.g.,]{KochanekSchechter:2004}.

\subsection{Inference from imaging data}\label{sec7:imaging_data}
High-resolution imaging of gravitational lenses with constraints from hundreds/thousands of high signal-to-noise surface brightness pixels is able to measure accurate astrometry of multiple images and capture the detailed distorted images of extended source structure. This information is crucial to capture the relative lensing deflection and to achieve a precise determination of the relative Fermat potential between multiple images of the time-variable source. Modeling the imaging data on the pixel level has become the standard over the last two decades.
In this section, we discuss the necessary aspects of the mass distribution that imaging data can constrain, apart from the remaining degeneracies.

To derive constraints on the mass of the gravitational lens and its deflection field from imaging data, models of the imaging data with different deflection fields are compared to the data in a Bayesian way on the likelihood level of individual pixels. Besides a description of the deflection field, all light emission components have to be described, containing the light emission of the source and the deflector.
All components that affect the imaging data need to be modeled and accounted for, in particular around the region impacted by lensing features. Required modeled components include, but are not limited to, the extended source component of host galaxy of the time-variable source, the image positions of the time-variable source and its resulting approximate point-like flux emission, the surface brightness of the deflector galaxy, differential dust extinction caused by the deflector galaxy on the background source, and any other sources of surface brightness, such as satellite galaxies. In addition, instrument effects, such as the point spread function (PSF), instrumental noise and shot noise, and pixelization, as well as potential data reduction artifacts need to be accurately taken into account in the modeling and comparison with the data.

The lensing effect distorting an extended surface brightness $S(\boldsymbol{\beta})$, such as from the extended host galaxy, can be computed with a method known as `backwards ray-tracing'.
Making use of the fact that surface brightness is conserved through lensing, the surface brightness at a position in the image $I(\boldsymbol{\theta})$ can be computed by the surface brightness in the source plane associated with the corresponding coordinate $\boldsymbol{\beta}(\boldsymbol{\theta})$ as
$I(\boldsymbol{\theta}) = S(\boldsymbol{\beta}(\boldsymbol{\theta}))
$,
where $\boldsymbol{\beta}(\boldsymbol{\theta}) = \boldsymbol{\theta} - \boldsymbol{\alpha}(\boldsymbol{\theta})$ is the `ray tracing' term given by the lens equation (Eqn.~\ref{eqn:lens_equation}).

For unresolved point-like images, the backwards ray-tracing is numerically inefficient. To guarantee that multiple images precisely come from the same location in the source plane within the astrometric requirements for an accurate time-delay prediction, the lens equation has to be solved for the point source constraints within the astrometric precision, or alternatively, solutions not satisfying the astrometric requirement \citep[e.g.,][]{Birrer:2019astrometry} need to be discarded.

Given a lens model with parameters $\boldsymbol{\xi}_{\rm mass}$ (which includes all aspects of the deflection field, including line-of-sight structure and nearby perturbers) and surface brightness model with parameters $\boldsymbol{\xi}_{\rm light}$, a model of the imaging data can be constructed, $\boldsymbol{d}_{\rm model}$.
The full process of simulating a modeled image can be cast as a consecutive application of operators as follows:
starting with the surface brightness operator $\mathcal{S}$, the lensing operator $\mathcal{L}$ is applied on the lensed source, followed by a PSF convolution operation $\mathcal{C}$, and finally an operator $\mathcal{P}$ matching the pixel resolution of the data, formally an integral of the convolved surface brightness over the size of a pixel.
With this notation and $\odot$ denoting the consecutive application of operators from left to right, we can write the generation of modeled data as
\begin{equation}
    \boldsymbol{d}_{\rm model} =  \mathcal{P} \odot \mathcal{C} \odot \left[\mathcal{L}(\boldsymbol{\xi}_{\rm lens}) \odot \mathcal{S}_{\rm source}(\boldsymbol{\xi}_{\rm light}) + \mathcal{S}_{\rm lens}(\boldsymbol{\xi}_{\rm light})\right].
\end{equation}

The Bayesian analysis to constrain the lens model is performed on the pixel-level likelihood of the imaging data. The likelihood is computed at the individual pixel level accounting for the noise properties from background and other noise properties, such as read-out, as well as the Poisson contribution from the sources. In the Gaussian limit the imaging likelihood is given by
\begin{multline}
  p(\mathcal{D}_{\rm img} \mid \boldsymbol{\xi}_{\rm mass}, \boldsymbol{\xi}_{\rm light}) \\
   = \frac{\exp \left[-\frac{1}{2}\left(\boldsymbol{d}_{\rm data} - \boldsymbol{d}_{\rm model}\right)^{\rm T} \boldsymbol{\Sigma}_{\rm pixel}^{-1}\left(\boldsymbol{d}_{\rm data} - \boldsymbol{d}_{\rm model}\right)\right]}{\sqrt{(2 \pi)^k {\rm det}(\boldsymbol{\Sigma}_{\rm pixel})}},
\end{multline}
where $k$ is the number of pixels used in the likelihood and $\boldsymbol{\Sigma}_{\rm pixel}$ is the error covariance matrix. We also note that for the flux noise, the error covariance matrix $\boldsymbol{\Sigma}_{\rm pixel}$ is a function of the brightness of the model $\boldsymbol{d}_{\rm model}$ and hence not independent of the model prediction. Current analyses assume uncorrelated noise properties in the individual pixels and the covariance matrix becomes diagonal.

The primary target of an imaging analysis is to retrieve the lens model parameter posteriors marginalized over other model parameters, in particular the surface brightness and regularization parameters as

\begin{equation}
    p(\boldsymbol{\xi}_{\rm mass} \mid \mathcal{D}_{\rm img}) = \int p(\mathcal{D}_{\rm img} \mid \boldsymbol{\xi}_{\rm mass}, \boldsymbol{\xi}_{\rm light}) p(\boldsymbol{\xi}_{\rm mass}, \boldsymbol{\xi}_{\rm light}) d \boldsymbol{\xi}_{\rm light},
\end{equation}
where $p(\boldsymbol{\xi}_{\rm mass}, \boldsymbol{\xi}_{\rm light})$ denotes the prior on the lens and light model parameters.

To jointly marginalize over an unknown yet possibly complex source morphology, different techniques have been developed. Such techniques include regularized pixelated source reconstruction \citep[e.g.,][]{WarrenDye:2003, Treu:2004, Koopmans:2005, Suyu:2006, Suyu:2009}, a set of basis functions, such as shapelets \citep[e.g.,][]{Birrer:2015, Birrer:lenstronomy} or wavelets \citep{Joseph:2019, Galan:2021}, or simply parameterized surface brightness profiles, such as S\'ersic profiles.
The methods mentioned above have in common that their surface brightness amplitude components create all a linear response on the pixel values of the data.
The optimization of the often numerous linear coefficients to provide the maximum likelihood of the data given a proposed model for the other parameters (such as surface brightness shape parameters and lensing parameters) can be cast as a linear problem with a solution obtained by a weighted least square minimization. 
The Gaussian covariance matrix of  the linear weighted least square minimization can be used to analytically marginalize over the prior of the linear coefficients \citep[e.g.,][]{ Suyu:2006, Vegetti:2009, Birrer:2015}\footnote{This is not a statement about the validity of the form of the prior of the linear surface brightness coefficients.}.

The joint sampling of lens and light model parameters to infer the lens model posterior distribution is then a semi-linear process. While the amplitudes of the light model coefficients can be solved linearly, the remaining parameters, including those pertaining to the lens mass model, and other shape-related surface brightness parameters have to be sampled non-linearly.

Often it is not clear at the beginning of an investigation what the level of complexity in the model is required to describe the data and to guarantee an accurate modeling.
Current procedures are to start with a simple model and subsequently increase the complexity in the different model components until a satisfactory fit is achieved. Current criteria for a goodness of fit in use are the Bayesian Information Criteria (BIC) \citep{Birrer:2019} and the Bayesian Evidence \citep{Shajib:2020}.

\begin{figure*}[htbp]
\centering
    \includegraphics[width=0.8\textwidth]{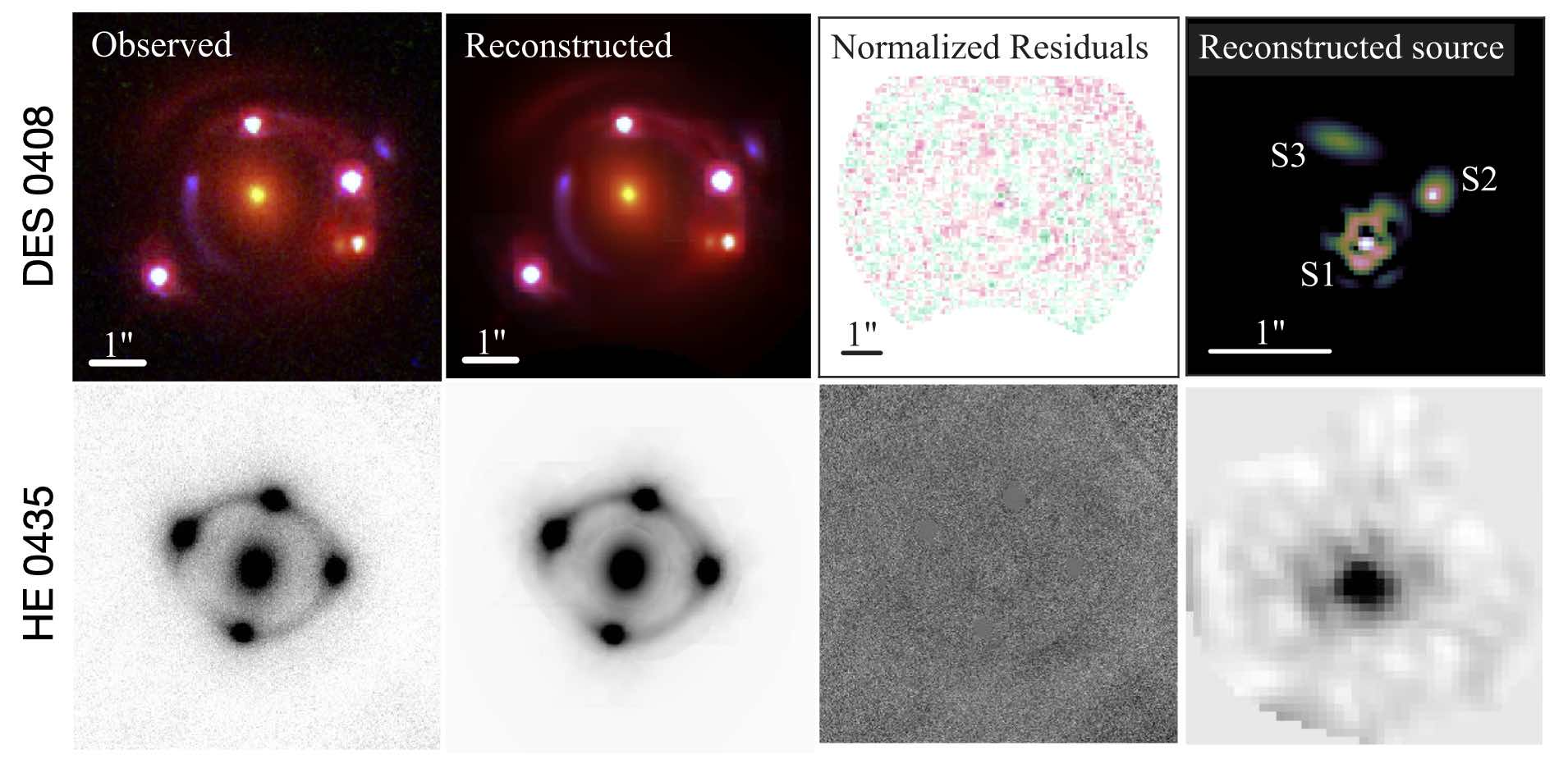}
    \caption{Illustration of imaging modeling for two lenses. From left to right: Imaging data, the reconstructed model, the reduced residuals $(\boldsymbol{d}_{\rm data}-\boldsymbol{d}_{\rm model})/\sigma$, reconstructed source. Top row: HST data and model for DES J0408$-$5354 in three bands, from \cite{Shajib:2020}, with shapelet and parameterized source reconstruction using the modeling software \textsc{lenstronomy}. Bottom row: Keck adaptive optics imaging and modeling of HE0435$-$1223, from \cite{Chen:2019}, with pixelated source reconstruction using the modeling software \textsc{GLEE}.}
    \label{fig:td_modeling}
\end{figure*}

\begin{figure*}[htbp]
\centering
    \includegraphics[width=0.8\textwidth]{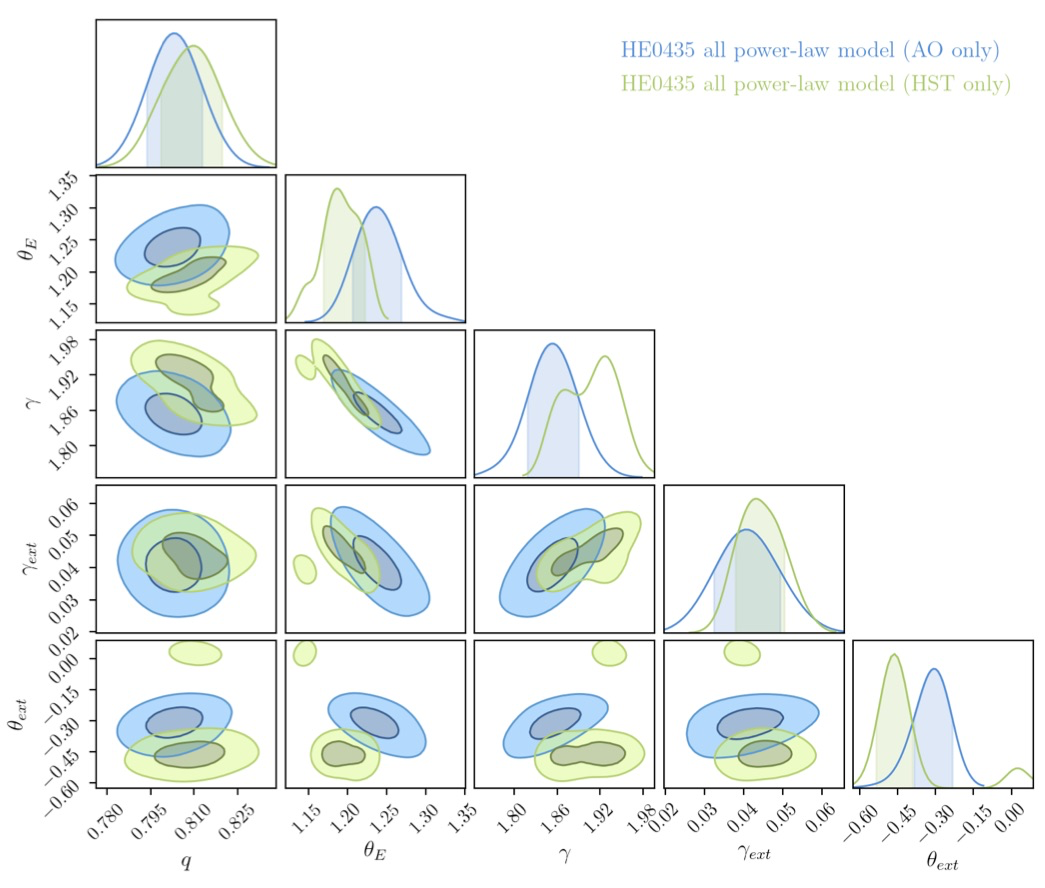}
    \caption{Key lens model parameter posterior from the fit to imaging data (\textit{HST} and Keck adaptive optics). $q$ is the semi-minor to semi-major axis ratio in the projected mass density profile, $\theta_{\rm E}$ is the Einstein radius, $\gamma$ the three-dimensional radial power-law density slope, $\gamma_{\rm ext}$ is the external shear strength, and $\theta_{\rm ext}$ is the shear angle. Figure adopted from \cite{Chen:2019}.}
    \label{fig:lensmodel_posterior}
\end{figure*}

% Comment on use of multiple filters and how the lens can be modeled simultaneously across bands.
Imaging modeling is primarily performed on high resolution space based \textit{Hubble Space Telescope} (\textit{HST}) or ground-based adaptive optics \citep[AO;][]{Chen:2016, Chen:2019, Chen:2021} imaging.
Figure~\ref{fig:td_modeling} illustrates, as an example, the imaging data and models for two quadruply imaged lensed quasars, originally presented by \citet{Shajib:2020} and \citet{Chen:2019}. Figure~\ref{fig:lensmodel_posterior} presents the key lens model posteriors from the imaging modeling fit of the lens HE0435$-$1223 by \cite{Chen:2019} for both, \textit{HST} and AO imaging, for a power-law elliptical mass distribution with external shear contribution.
To enhance the signal in the data set, and to distinguish deflector and source light components, the lens modeling is often performed simultaneously with multiple filters and combined on the likelihood level.
With multiple filters, a better differentiation between source and deflector light can be drawn, with the deflector often being bright in the infrared channels, and the lensed source often dominant in the optical and ultraviolet channels. The infrared channels are often brighter, and hence contain more signal, while the optical channels can resolve smaller angular scales with more prominant source morphologies.
The modeling across bands is performed with an identical lens model, while the surface brightness solutions are flexible to change (i.e., independent linear optimization or even different surface brightness components). Very high relative astrometric solutions of order $\sim 1$mas across bands are required for the modeling. Current modeling fits the relative astrometric solutions across bands in the modeling process \citep[e.g.,][]{Shajib:2020}. 

The PSF needs to be characterized very accurately, both to provide a high astrometric precision of the images of the time-variable sources \citep{Birrer:2019astrometry, Chen:2021astrometry} and for the detailed modeling of the extended source structure without spurious signal of bright quasar images. Current methods to obtain a precise PSF model contain an iterative procedure during the model fitting process to extract improved constraints of the PSF from the data itself \citep[e.g.,][]{Chen:2016, Birrer:2016}.

The lens model and Fermat potential posteriors are marginalized over a set of systematic effects and modifications in the choice of the source reconstruction and other modeling choices.

\subsection{Mass profile assumptions of the main deflector}\label{sec7:mass_profile}

\begin{figure*}[htbp]
\centering
    \includegraphics[width=0.95\textwidth]{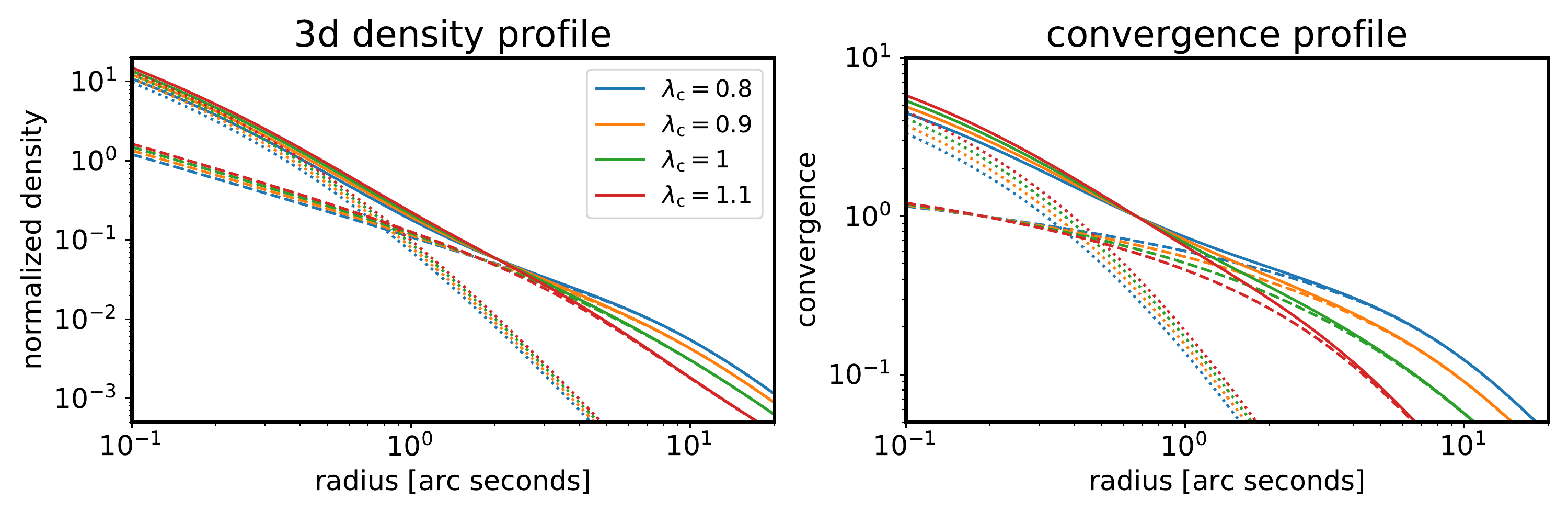}
    \caption{Illustration of a composite profile consisting of a stellar component (Hernquist profile, dotted lines) and a dark matter component (NFW + cored component with $\lambda_c$ acting as an approximate MST \citep[from][Eqns. \ref{eqn:cored_density}, \ref{eqn:mst_approx}]{Blum:2020}, dashed lines) which transform according to an approximate MST (joint as solid lines). The stellar component gets rescaled by the MST while the cored component transforms the dark matter component. Physically, the profiles of each color differ by a 10\% different mass-to-light ratio combined with a slightly more extended or contracted dark matter profile also on the 10\% level. \textbf{Left:} profile components in three dimensions. \textbf{Right:} profile components in projection. Each profile provides a 10\% difference in the predicted time delay, and hence $H_0$ inference. The transforms presented here cannot be distinguished by imaging data alone and require i.e., stellar kinematics constraints.
    Figure from \cite{Birrer:tdcosmoiv}.}
    \label{fig:mst_composite}
\end{figure*}

Resolved multiply imaged structure is an exquisite data product to provide constraints on the relative deflection field (see Section~\ref{sec7:imaging_data}). 
Imaging data, in the absence of the knowledge of the unlensed apparent source size or brightness, is not able to break the MST and its generalization, the source position transformation \citep[SPT;][]{Schneider:2014}.
The quantity that is invariant under the MST in the radial direction and hence can be constrained by imaging data is \citep{Kochanek:2002, Sonnenfeld:2018, Kochanek:2020, Birrer:2021curvedarcs}
\begin{equation}\label{eqn:rad_constraint}
	\xi_{\rm rad} \equiv  \frac{\theta_{\rm E} \alpha_{\rm E}^{\prime \prime}}{1-\alpha_{\rm E}^{\prime}} \propto \frac{\theta_{\rm E} \alpha^{\prime\prime}_{\rm E} }{1 - \kappa_{\rm E}},
\end{equation}
where $\alpha^{\prime}_{\rm E}$ is the derivative and $\alpha^{\prime\prime}_{\rm E}$ is the second derivative of the deflection angle at the Einstein radius $\theta_{\rm E}$, respectively, and $\kappa_{\rm E}$ is the convergence at $\theta_E$.
On azimuthal invariances and observable lensing features, we refer to \cite{Birrer:2021curvedarcs} and references therein.
Constraining a global mass density profile based on imaging data alone requires assumptions on the radial profile. For example, when imposing the assumption that the mass density profile follows a single power law, the power-law slope $\gamma_{\rm pl}$ has a direct correspondence to $\xi_{\rm rad}$ (Eqn. \ref{eqn:rad_constraint}) with $\xi_{\rm rad} = \gamma_{\rm pl} - 2$ and a precise (few percent uncertainty) inference on the Fermat potential is possible.
Relaxing the assumption on the shape of the density profile leads to significantly widened constraints. When allowing for a full mass-sheet degree of freedom, the Fermat potential is fully degenerate (i.e. $\xi_{\rm rad}$ remains unchanged by a MST).
A pure mass sheet is unphysical as no localized three-dimensional density profile can describe it. However, approximate parameterizations can be found that can be expressed as a three-dimensional density profile and are indistinguishable based on imaging data \citep{Schneider:2013, Blum:2020, Birrer:tdcosmoiv, Yildirim:2021}.
For example, \cite{Blum:2020} introduced the cored density profile which has a three-dimensional density distribution
\begin{equation}
    \rho_c(r) = \frac{2}{\pi} \Sigma_{c} R_c^3 \left(R_c^2 + r^2 \right)^{-2}
\end{equation}
with the convergence profile as
\begin{equation} \label{eqn:cored_density}
    \kappa_c(\theta) = \left(1 + \frac{\theta^2}{\theta_c^2} \right)^{-3/2}.
\end{equation}
The approximate MST can then be written as
\begin{equation} \label{eqn:mst_approx}
    \kappa_{\lambda_c}(\theta) = \lambda_c \kappa(\theta) + (1-\lambda_c) \kappa_c(\theta).
\end{equation}
Figure~\ref{fig:mst_composite} illustrates an example of how an approximate MST can be physically interpreted when applied to a composite profile described with a stellar and a dark matter component.

There are also possibilities in deviations in the mass density profiles that do not directly mimic an MST. Any radial mass profile that satisfies the same constraints on $\xi_{\rm rad}$ (Eqn. \ref{eqn:rad_constraint}) provides an equally good fit to the data\footnote{$\xi_{\rm rad}$ is the first-order Taylor expansion term affecting the observed radial distortions. The quantity is well defined for an azimuthally symmetric lens. The generalization of the relevant radial quantity for elliptical mass models needs to be investigated.}.
Azimuthal assumptions of the mass density profile do also matter in the interpretation of the radial components
\citep{Birrer:2021curvedarcs, Kochanek:2021}
and assumptions on disky and boxiness, ellipticity gradients and isodensity twists of the density profile may also impact the Fermat potential differences \citep{vdVyvere:2022, vandeVyvere:2022b, Gomer2020, Gomer2021}.

From a physics point of view, the matter distribution of the main deflector is made of stellar mass, gas, and dark matter, where the stellar mass is dominating the inner-most parts. The dark matter fraction within the Einstein radius is about $\sim 10-60\%$ \citep[e.g.,][]{Auger10b, Ferreras2005}.
Invisible substructure in the lens and along the LOS can also perturb the Fermat potential \citep[e.g.,][]{Oguri:2007, Keeton:2009}. 
\cite{Gilman:2020} showed that omitting dark substructure does not bias inferences of $H_0$. However, perturbations from substructure contribute an additional source of random uncertainty in the inferred value of $H_0$ ranging from 0.7 - 2.4\% depending on the redshift and image configuration.
We also highlight that the lensing mass and convergence only accounts for the mass over-density in regard to the cosmological background density.

Different approaches running with different assumptions have been taken in the literature to describe the deflector lensing potential. 
Among the assumptions being used are single power-law mass profiles, composite models involving a mass-follows-light component with a separate component describing the dark matter profile, free-form pixelated mass profiles \citep[e.g.,][]{Saha2004, Coles:2014, Denzel:2021}, pixelated lensing potential corrections \citep[e.g.,][]{Suyu:2009}, or an explicit internal mass profile MST component \citep{Blum:2020, Birrer:tdcosmoiv}.

There are multiple considerations in the specific choice of an investigation. On one side, there are physical considerations. What basic assumptions in the modeling are justified?  What priors to chose in the Bayesian modeling?
Then there are also practical considerations. What aspects of the model can be constrained by the data? Is it feasible to perform a posterior inference in a finite amount of time with given resources?

Among the simplest models employed is the single power-law profile. It has a one-to-one relation to the radial quantity described in Equation~\ref{eqn:rad_constraint} and breaks the MST. A power-law elliptical mass profile is an efficient parameterization to describe the first order radial and azimuthal features in strong lensing imaging data. Composite models do relate to certain physical assumptions of mass follows light and assert a stiffness in the profile that implicitly also break the MST \footnote{The breaking of the MST for composite models is dependent on imposed mass and concentration priors of the dark matter profile, as well as mass-to-light gradients or the absence of it.}.

An explicit parameterization of the MST in the model denies any prior or assumptions to break the MST and is maximally agnostic to the MST with minimal added parameter degrees.

On the high-complexity end of lens models are free-form methods, such as pixelated mass distributions \citep{Saha2004, Coles:2014}. Free-form models come with very few restrictions on the lens mass distribution and offer a different modeling strategy compared to the simply parametrized approaches. The ensemble of models allowed by the data can be interpreted as the model posterior distribution, with the regularization scheme proposing models without data constraints being the prior.

Increased flexibility in the parameterization better guarantees that the underlying truth in the mass distribution, and in particular the prediction of the Fermat potential entering the time delays, can be represented by the model.
On the other hand, increased flexibility in the model at fixed data constraining power increases the uncertainty in the posterior-predictive model. Less constraining posteriors put inevitably more weight and reliance on the priors applied, whether they are explicitly in a parameterized form, or implicit within an over-parameterized, free-form approach. No matter what choices are being made in the modeling of lenses, mitigating the dependence on the explicit or implicit priors becomes important when combining a set of multiple lenses, as we will discuss in Section~\ref{sec7:population_inference}.

We discuss additional data sets that can constrain the lens mass profile in Section~\ref{sec7:external_data}.

\subsection{Non-lensing observables}\label{sec7:external_data}

\begin{figure}[htbp]
\centering
    \includegraphics[width=0.45\textwidth]{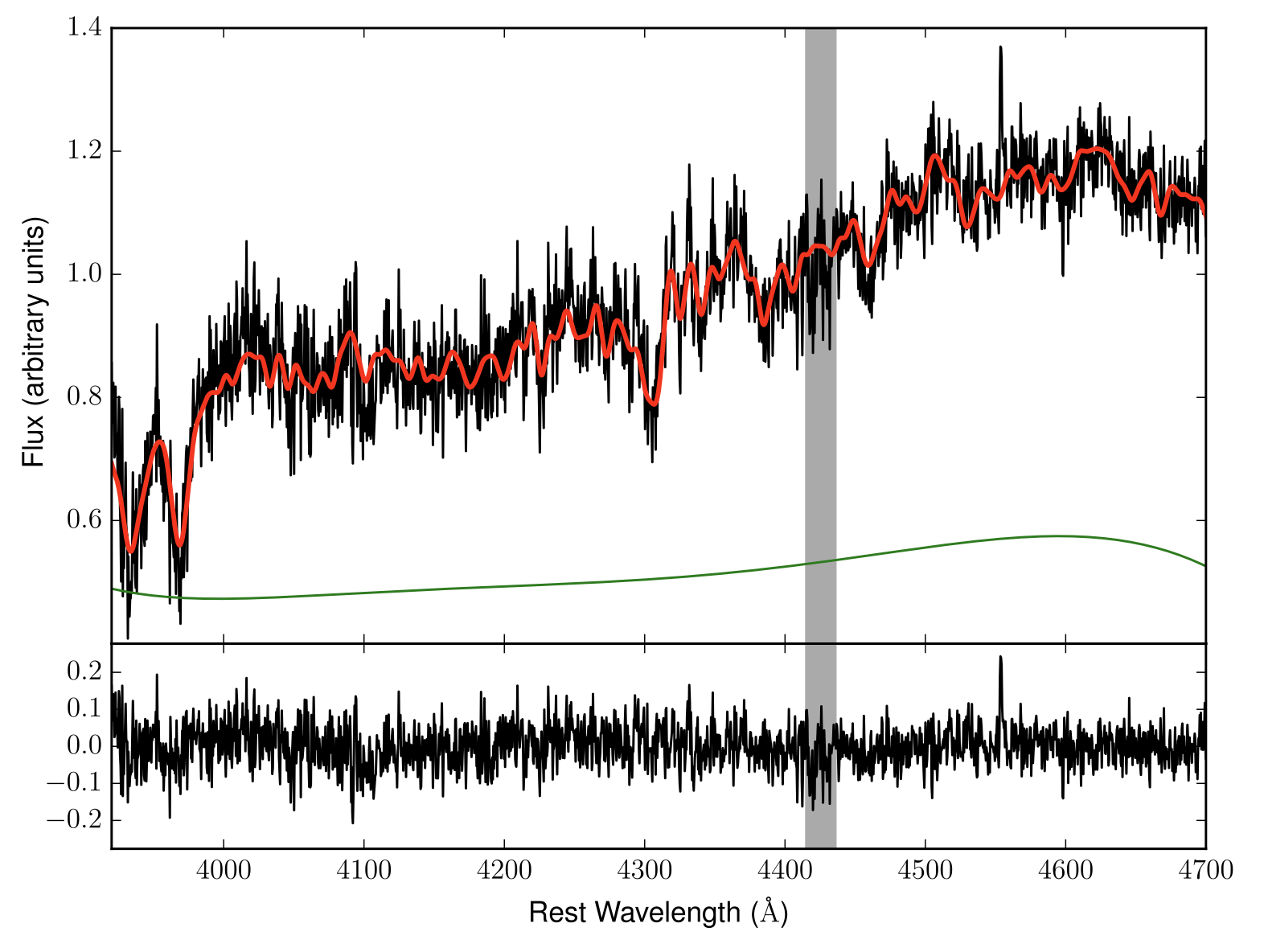}
    \caption{\textbf{Top}: Keck/LRIS spectrum of HE0435$-$1223 with the best-fitting model overplotted in red and a polynomial continuum, which accounts for contamination from the lensed QSO images and template mismatch, shown in green. The measurement results in an integrated velocity dispersion $\sigma_{\rm v} = 222\pm15$ km s$^{-1}$, including systematic uncertainties due to the templates used, the region of the spectrum that was fitted, and the order of the polynomial continuum. The grey vertical band represents a wavelength range that is excluded from the fit due to the presence of a strong Mg II absorption system. Bottom: Residuals from the best fit. 
    Figure adopted from \cite{Wong:2017}.}
    \label{fig:keck_spectrum}
\end{figure}

The currently used primary observation to break the MSD is stellar kinematics from the deflector galaxy \citep{Treu:2002, Koopmans:2003, Koopmans:2004}.
The kinematics of stars, in particular their velocity dispersion, is a direct and lensing-independent tracer of the three-dimensional gravitational potential.
The kinematic measurement is performed by targeting stellar absorption lines and measuring their width with spectrographs. Figure~\ref{fig:keck_spectrum} shows a Keck/LRIS spectrum of HE0435$-$1223.

The line-of-sight stellar velocity dispersion is an integrated quantity of the radial and tangential components of the velocity dispersion projected along the line of sight. The orbital anisotropy, i.e., the ratio of the radial and tangential velocity dispersion components, is unknown \textit{a priori} and thus introduces a degeneracy in the predicted line-of-sight velocity dispersion corresponding to the same 3D mass profile. This degeneracy is known as the mass--anisotropy degeneracy. Typically, a prior on the anisotropy profile, e.g., isotropic or Osipkov--Merritt \citep{Osipkov79, Merritt85, Merritt85b}, is assumed. The Osipkov--Merritt profile allows the anisotropy to be isotropic near the center and gradually more radial farther away from the center, which is motivated by the observed properties of the stellar orbits in local elliptical galaxies. The isotropic profile is thus a special case of the Osipkov--Merritt profile. The transition scale radius $r_{\rm ani}$ from isotropic to radial orbits in the Osipkov--Merritt profile is \textit{a priori} unknown, which directly manifests in the mass--anisotropy degeneracy. Thus, the prior on $r_{\rm ani}$ has significant impact on the kinematics prediction \citep[e.g.,][]{Shajib18, Birrer:tdcosmoiv}. We note that there are many other forms of the radial anisotropy distribution and the specific choice of functional model used might impact the results, as well as what priors are adopted. One way to mitigate this degeneracy is to obtain spatially resolved measurement of the velocity dispersion instead of an unresolved (or, integrated) velocity dispersion \citep{Shajib18, Yildirim:2020, Yildirim:2021, BirrerTreu:2021}.
%end text contributed by AJS

Other proposed observations and analyses methods that can break the MST and provide the necessary constraints on the mass density profile are standardizable magnifications \citep[e.g.,][]{Kolatt:1998, Oguri:2003, Foxley-Marrable:2018, Birrer:2021glSNe}, lens population statistics of appearances and asymmetry in the multiple images \citep[e.g.,][]{Sonnenfeld:2021a, Sonnenfeld:2021b}, and galaxy-galaxy weak gravitational lensing \citep{Khadka:2024}.

We emphasize that these non-lensing observations are primarily sensitive to the total MST, the combination of LOS and internal mass density profile degeneracies (Eqn.~\ref{eqn:mst_combined}). Decoupling of the different projected effects in the lensing potential is not necessary to perform an accurate cosmographic inference since the time-delay prediction only requires the combined product.
However, when combining different lenses with potentially different selections, the priors and assumptions imposed in either of the two components impacting the MST can become important.

%%%%%%%%

\section{Estimating line-of-sight contributions}\label{sec7:los}

Strong lensing requires a high projected mass density. Strong lenses are hence biased toward more massive galaxies, which are biased toward overdense environments.
The contribution of the mass density fluctuations along the line of sight to the lensed source is generally of order few percent, and commonly lower than 10\% of the total convergence of the lens. While this may appear to be small, it is not negligible when it comes to estimating the Hubble parameter to percent accuracy. A constant effective contribution of a few percents caused by the line-of-sight is equivalent to an external mass-sheet $\kappa_{\rm ext}$  (Sect.~\ref{sec7:distance_measure}). 

The exact impact of the line-of-sight objects depends on whether the dominant-lens approximation is valid, in which case the critical density of the line-of-sight objects is very small compared to the main deflector critical density, and on whether the tidal regime is applicable, which happens when the perturber's gravitational field is small compared to the changes of the deflection $\alpha(\theta)$ \citep[e.g.][]{McCully:2014, Birrer:2017los, Fleury:2021los}. 
When one of those approximation is invalid, an explicit treatment is needed, requiring potentially to solve the multi-plane lens equation \citep[see e.g., ][]{McCully:2014, Wong:2020, Shajib:2020, Li:2021}. We refer to \chapintro~ for a detailed discussion on the multi-plane gravitational lensing formalism. Conversely, when line-of-sight objects can be treated as small perturbations that only introduce convergence that is constant over the extent of the lensed system, a statistical treatment is sufficient. In practice, a hybrid scheme needs to be followed most of the time, including explicitly modelling those objects that modify differently the Fermat potential for each lensed images, and calculating the statistical contribution of the other objects that shift the Fermat potential in linear order. 

From an information perspective, there is only limited direct data available of the total matter distribution on the universe at the scales relevant ton constrain $\kappa_{\rm ext}$. Hence, any method relies on some assumptions on how mass traces light. These assumptions are well motivated by large scale structure probes, but are only validated in a statistical way.

The following subsections present the various methods that have been considered to estimate $\kappa_{\rm ext}$. Section~\ref{chap3:los_direct_modeling} presents a direct modeling, Section~\ref{chap3:los_number_counts} presents galaxy number counts statistics, Section~\ref{chap3:los_wl} weak lensing measurements, and Section~\ref{chap3:los_hybrid} a hybrid approach.

\subsection{Direct modeling} \label{chap3:los_direct_modeling}

\begin{figure*}[htbp]
\centering
    \includegraphics[width=0.8\textwidth]{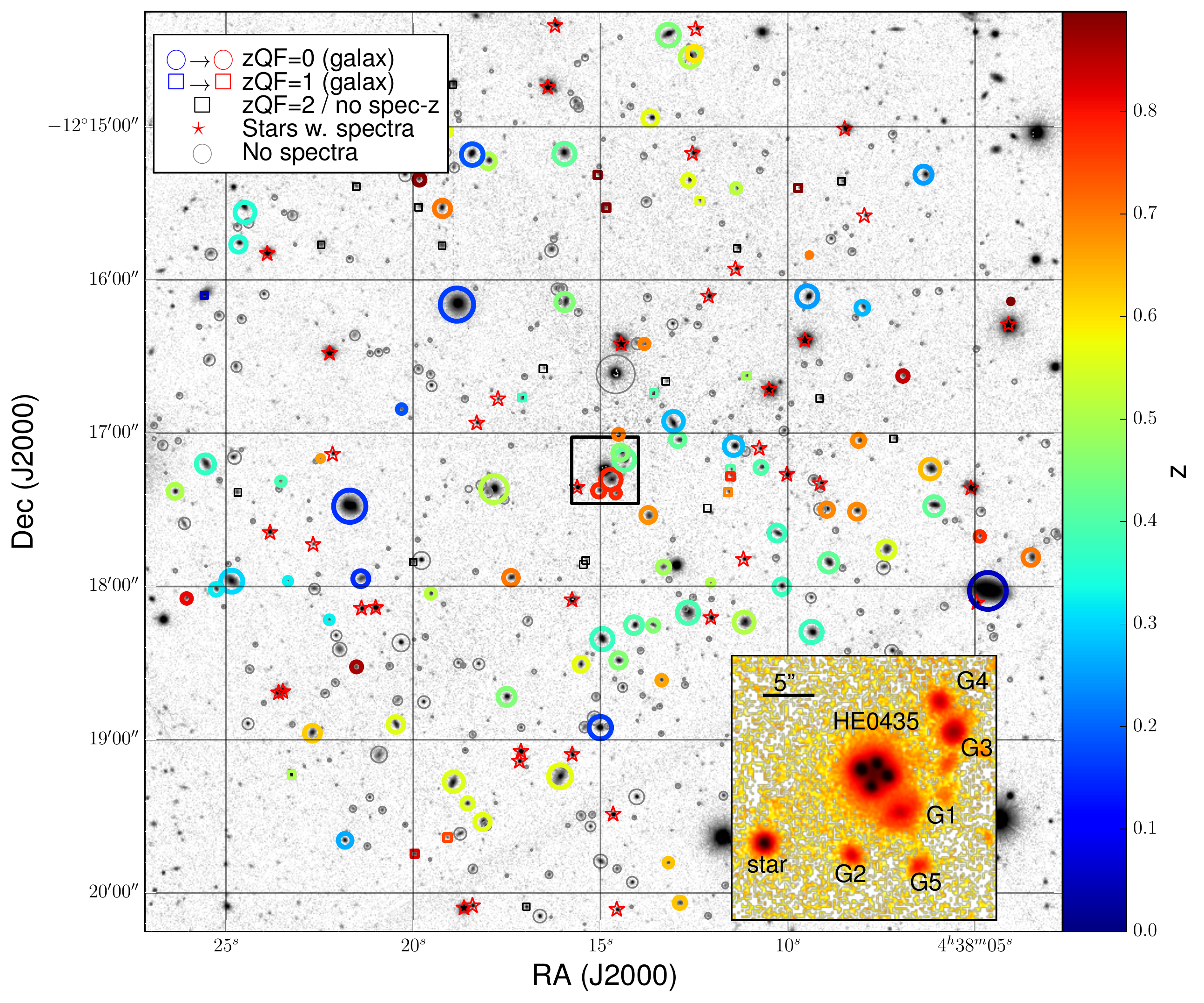}
    \caption{Example of the result of the spectroscopic characterization of the field of view of the gravitational lens HE0435$-$1223. The redshifts of the objects measured in the field of view are used to identify which objects (or galaxy groups) needs to be explicitly included in the macro-model. The number counts analysis uses the galaxies located at projected distances of $<$45$^{\prime\prime}$ and $<$120$^{\prime\prime}$ from the lens to estimate $P(\kappa_{\rm ext})$. Courtesy of  \citet{bib7:Sluse2017} }
    \label{fig:los}
\end{figure*}

The most direct approach is to collect the positions, redshifts, stellar masses and potentially even velocity dispersion measurement of the galaxies located in the field of view towards the lens and explicitly model the matter distribution of all relevant objects. A complete direct reconstruction is near-impossible.
A simple approach that has been developed to estimate $\kappa_{\rm ext}$ consists in identifying which galaxies form massive galaxy groups that contribute the most significant impact along the line of sight \citep[e.g.][]{bib7:Fassnacht2002, bib7:Momcheva2006, Fassnacht2006, bib7:Sluse2017}. An estimate of $\kappa_{\rm ext}$ may then be derived by fitting analytical mass density profiles on those groups \citep{bib7:Auger2007, bib7:Wilson2017}. This approach generally yields estimates of $\kappa_{\rm ext}$ typically uncertain to a factor 2-4 depending on the specific assumptions one may reasonably make on the group mass density (e.g., halo associated to each individual galaxies, a common halo for all the systems), but also sometimes with low precision due to the uncertainties associated to the group identification (fields of view are never complete and group finders have their own biases). To overcome this problem, \cite{bib7:Collett2013} have proposed a simple halo model prescription to reproduce the mass along the line-of-sight from a photometric catalogue of galaxies. The convergence $\kappa_{\rm h}$ from each halo has then been calibrated against $\kext$ derived from ray-tracing estimate through numerical simulations. This method does not account explicitly for the convergence due to dark structures and divergence due to voids, but those effects are included statistically owing to the calibration of $\kappa_{\rm h}$ against $\kext$.

\subsection{Number counts} \label{chap3:los_number_counts}

\begin{figure}[htbp]
\centering
    \includegraphics[width=0.5\textwidth]{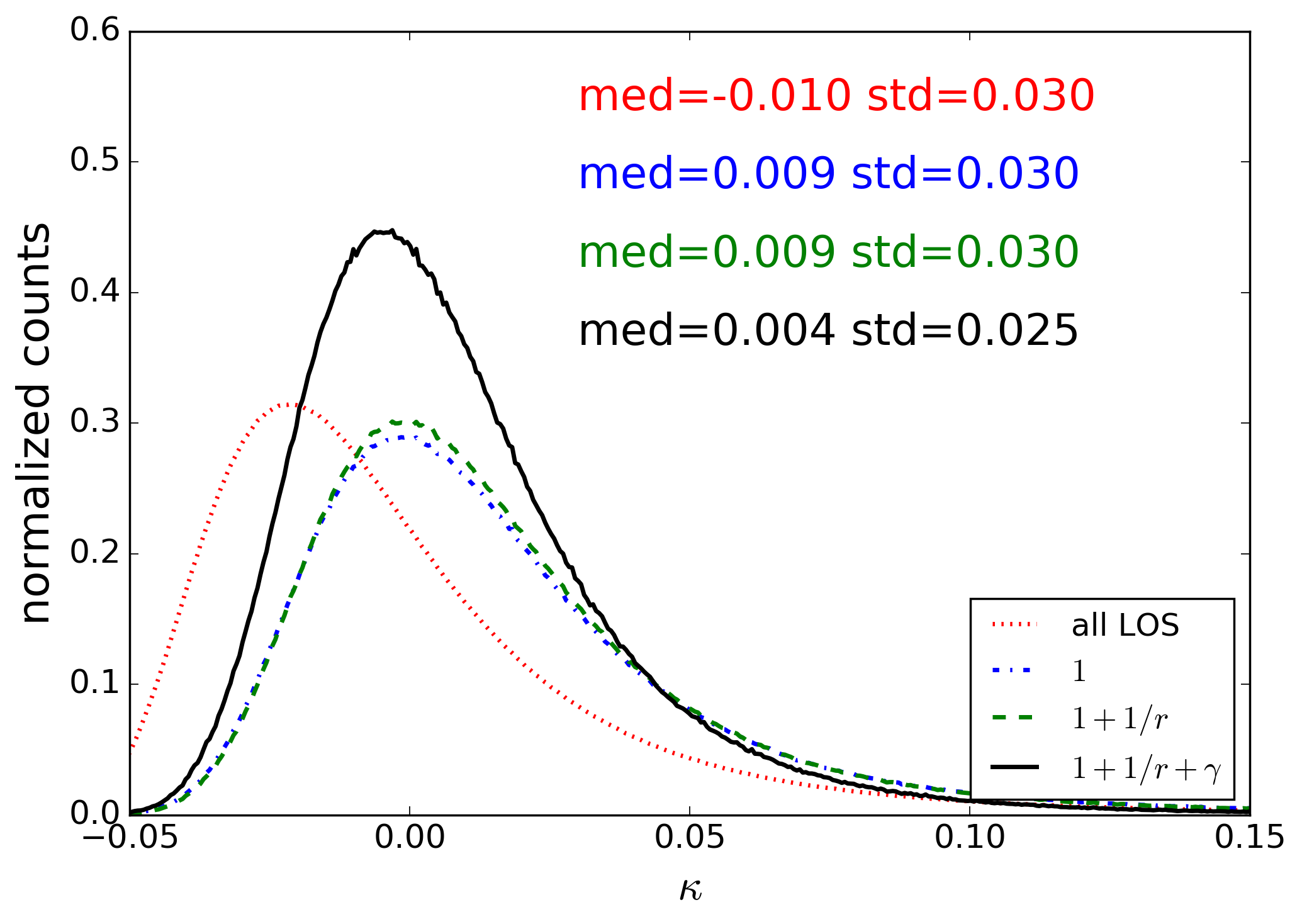}
    \caption{Probability distribution $P(\kappa_{\rm ext})$ for HE0435$-$1223 and an aperture radius of 45$^{\prime\prime}$. The different curves correspond to all lines of sights (dotted red), considering only lines of sights with the same overdensity as the data (dash-dotted blue), using a weighting inversely proportional to the distance (dashed green) and the additional constraint from the shear (solid black). Courtesy of  \citet{Rusu:2017}}
    \label{fig:kappa_ext}
\end{figure}

An alternative to the direct modeling of the line of sight consists of measuring the galaxy number density in the vicinity of the lens, describe it in a summary statistic, and comparing it to reference fields. This procedure will determine whether the LOS is over- or under-dense compared to average  \citep[][]{Suyu:2010, bib7:Fassnacht2011, Greene:2013, Rusu:2017, Wells2023}. First, a detailed characterization of the line of sight towards the lens is required, using deep imaging data, complemented by spectroscopic data (see Fig.~\ref{fig:los} for an illustration). Similar LOS are then searched for in large volume, and high resolution  numerical simulations. The surface mass density of matter along these line of sight being calculated using a ray-tracing technique \citep{bib7:Hilbert2007, bib7:Hilbert2008, bib7:Hilbert2009}, it is then possible to derive a probability distribution of external convergence compatible with the observed lens matching the summary statistics.

In practice, summary statistics that deviate from pure number counts can be a better informed statistics of the underlying over- or under- density. For example, if $N_{\rm{gal}}$ galaxies are observed in the field of view of a lens, one can calculate a weighted number counts $W_q = \sum_{i=1}^{N_{\rm{gal}}} q_i$ with $q$ being a particular type of weight, such as the inverse of the distance to the lens, i.e. $1/r$. To calculate a weighted density of galaxies, $\zeta_q$, it is necessary to perform the same measurement over an ensemble of control fields, such that for each control field $({\rm CF},\,j)$ one derives a density $\zeta_q^{j} = W_q / W_{q}^{{\rm CF},\,j}$. To avoid introducing any bias through this normalisation procedure, it is important to choose enough control fields but also ensure that those fields have characteristics that match closely those of the imaging data of the observed lens system. This allows one to account for sample variance and to assess that galaxy detection biases are similar for the lens and for the control fields. In particular, one should ascertain that the lens and control fields have similar depth, seeing, and pixel scale, the latter quantities being critical in the framework of source deblending and identification. It happens that some regions of the control fields and potentially the data around the lens target are masked due to saturation of stars, cosmic rays, or camera defects. To guarantee unbiased estimates of $\zeta_q$, it is important to apply the same mask to the weighted count of the lens field and the control field. A large variety of weighting schemes has been explored, some of them involving a proxy on some of the galaxy properties such as the redshift and the stellar mass. Those quantities are derived using a photometric redshift code, such as \texttt{LEPHARE} \citep{bib7:Ilbert2006}. This implies the availability of deep multi-band photometric data. Having a similar depth for the lens and comparison field is important to matching galaxy properties in photometric surveys. Also, similar set-up for magnitude measurements are required to minimize systematic errors caused by aperture and/or object deblending uncertainties. When spectroscopic redshift are also available, they may generally be preferred to the photometric ones. 

The conversion of the probability density distribution $p(\zeta_q)$ into $p(\kext)$ requires the use of numerical simulations for which $\kext$ has been derived through ray-tracing. Large simulation volumes are required to minimise the impact of sample variance and cosmic variance. The Millennium simulation \citep{bib7:Springel2005} being dark matter only, galaxy photometric properties are inpainted using physically motivated prescriptions. A common choice has been to use the semi-analytic model of \cite{bib7:Delucia2007}. Densities $\zeta_q$ can be estimated in those catalogues in similar way as for real data except that the reference fields is now the simulation catalog itself. As explained in \cite{Rusu:2017}, the probability $p(\kext \mid \mathbf{d})$, where $\mathbf{d}$ stands for the available data, is given by:
\begin{multline}
p(\kext \mid \mathbf{d})  = \int d \zeta_q \frac{p(\kext,\zeta_q) p(\zeta_q,\mathbf{d})}{p(\zeta_q) p(\mathbf{d})}  = \\ 
\int d \zeta_q p(\kext \mid \zeta_q) p(\zeta_q \mid \mathbf{d})   \ .
\end{multline}

\cite{Greene:2013} has shown that the precision on $\kappa_{\rm ext}$ is improved by a factor 2 when using a combination of weights, the two main ones being the standard number counts ($q=1$), and as the inverse of the projected distance to the lens ($q=1/r$). The addition of a weight based on the modeled amplitude of the shear, $\gamma_{\rm ext}$ is also often considered, such that the general expression of $p(\kappa_{\rm ext} \mid \mathbf{d})$ becomes: 
\begin{multline}
p(\kappa_{\rm ext} \mid \mathbf{d}) = \int d \zeta_1 d \zeta_{1/r} d \zeta_{q\neq1,1/r} d \zeta_{\gamma_{\rm ext}} p_\mathrm{MS}(\kappa_{\rm ext} \mid  \zeta_1, \zeta_{1/r}, ... \\  
... \zeta_{q\neq1,1/r}, \zeta_{\gamma_{\rm ext}}) p(\zeta_1, \zeta_{1/r}, \zeta_{q\neq1,1/r}, \zeta_{\gamma_{\rm ext}} \mid  \mathbf{d})  \ .
\end{multline}

The addition of a fourth weight, $q\neq1,1/r$ allows one to evaluate systematic errors involved by the specific choice of equally motivated weighting schemes and explore which combination of weight yields the best precision on $\kext$. The first application of this technique in the framework of time-delay cosmography has been presented in \cite{Suyu:2010}. Subsequent time-delay cosmography analyses from H0LICOW, STRIDES and TDCOSMO have used an approach that broadly follows the strategy outlined above \citep[see][for a more in-depth description of the method]{Greene:2013, Rusu:2017}, but proposing small variations and tests in terms of weighting schemes and choices of comparison fields \citep{Birrer:2019, bib7:Sluse2019, bib7:Buckley-Geer2020}. Figure ~\ref{fig:kappa_ext} displays $p(\kappa_{\rm ext})$ as derived with different weighting  schemes for the lens system HE0435$-$1223. 

A novel method for estimating $\kext$ has recently been proposed by \cite{Park:2021}. They replace the weighted number count scheme by a machine learning approach. Specifically, they have trained a Bayesian Graph Neural Network on LSST DESC DC2 sky survey \citep{bib7:LSST_DESC_DC2_2021} in order to derive a distribution of $\kappa_{ext}$ for arbitrary gravitational lens sight-line.

%There is a slight inconsistency/circularity in using a simulation with an assumed cosmology, but this is likely a small effect.  It should be discussed here for completeness.
The reliance on cosmological simulations and their cosmological assumptions poses a slight circularity in the inference when the very goal of time-delay cosmography is to test and challenge cosmological models. However, the line-of-sight correction term that is constrained relying on cosmological assumptions is perturbative, i.e., even if the actual cosmology resulted in a $\sim 10\%$ relative difference in the line of sight characteristics, it would be a sub 1\% change to the distance measurements since the expected effect is only a few percent.

\subsection{Weak Lensing} \label{chap3:los_wl}

Weak lensing, the linear shape distortion of background galaxies due to foreground structure, is a direct probe of the LOS structure. On linear scales, the cosmic shear measurements can be translated to convergence in a unique mapping \citep{Kaiser:1993}. Hence, this technique does neither rely on priors from numerical simulations nor of a galaxy-halo connection. However, there are also several drawbacks.
The angular scale of a weak lensing measurement is limited by the number density of lensed sources, and a high S/N measurement can only be achieved at scales of arc minutes. Thus, weak lensing is an excellent observable to quantify large scale cosmic density distributions but other smaller scale density perturbations down to the scales of arc seconds are not well captured. Another limitation is that the weak lensing source population is not at the same redshift as the strongly lensed source. One needs to translate the weak lensing convergence map to a different lensing kernel, which comes with additional statistical uncertainties \citep[e.g.,][]{Kuhn:2021}.

In the strong lensing context, for example,
\cite{Fischer:1997, Nakajima:2009, Fadely:2010} relied on the weak lensing effect produced by massive structures in the vicinity of the deflector. They constrained the external convergence by integrating the tangential weak gravitational shear in the area around the lens. 
More recently, \cite{Tihhonova:2018, Tihhonova:2020} applied the weak lensing techniques to the quadruply lensed quasar systems HE0435$-$1223 and B1608$+$656 and performed a convergence map reconstruction based on HST imaging. \cite{Kuhn:2021} performed a convergence map reconstruction of the COSMOS field at the position of discovered strong lenses.

\subsection{Hybrid framework} \label{chap3:los_hybrid}

Given the strengths and weaknesses of the direct modeling and summary statistics approaches, as well as weak lensing measurements, a hybrid approach can leverage the complementary methodologies. Summary statistics can be most effectively employed for objects that mostly cumulatively affect the lensing convergence while explicit modeling of LOS objects makes a difference for massive or very close by objects.
The specific decision of where to split the analysis between a statistical approach and explicit modeling is primarily impacted by two factors. The first is the accuracy in the deflection properties, both in terms of higher-order lensing distortions and the need for a multi-plane lensing approach. The second is the available information and the handling of priors in the absence of sufficient information.

A method to account accurately for the line-of-sight has been proposed by \cite{McCully:2014, McCully:2017}. It consists in a multi-plane lens equation where only the planes associated to important perturbing groups/clusters/galaxies are included. The other perturbers along the LOS are treated under the tidal approximation. In order to identify those objects, \cite{McCully:2017} proposes to use a threshold based on the value of the flexion-shift, i.e. $\Delta_3 x$ whose expression is given by:
\begin{equation}
\Delta_3 x = f(\beta) \, \times \frac{(\theta_{\rm E} \,\theta_{\rm E,p})^2}{\theta^3}, 
\end{equation}

\noindent where $\theta_{\rm E}$ and $\theta_{\rm E, p}$ are the Einstein radius of the main lens and of the perturber, and $\theta$ is the angular separation on the sky between the lens and the perturber. The function $f(\beta) = (1-\beta)^2$ if the perturber is behind the main lens, and $f(\beta) = 1$ if the galaxy is in the foreground. In that expression, $\beta$ is the pre-factor of the lens deflection in the multiplane lens equation: 

\begin{equation}
    \beta = \frac{D_{\rm{dp}} D_{\rm{os}} }{D_{\rm{op}} D_{\rm{ds}}}, 
\end{equation}

\noindent where $D_{ij} = D(z_i, z_j)$ are angular diameter distances between redshifts $z_i$ and $z_j$, corresponding to the observer ($\rm{o}$), deflector ($\rm{d}$), perturber ($\rm{p}$) and source ($\rm{s}$). Missing to accounting for a foreground perturber may have a stronger impact on the models than missing a background one. The reason is that the background perturber will have a multiplicative effect on the source position, while the deflection from the foreground pertuber enters the lens equation inside the argument of the deflection of the main lens galaxy. In other words, the foreground perturber modifies the coordinates of the lensed images positions compared to the main lens case. These non-linear effects require a multi-plane treatment to be properly accounted for. From a set of simulation of time-delay lens systems resembling real ones, and their subsequent modeling based on point-source image positions, \cite{McCully:2017} suggests that a value $\Delta_3 x < 10^{-4}$\, arcsec yields to a bias on $H_0$ of less than a percent. Since \cite{bib7:Sluse2017}, this prescription is used by the H0LICOW and TDCOSMO collaborations to select the objects that they explicitly include in the lens mass modelling.

\cite{Birrer:2017los, Kuhn:2021} combined the study of the environment using the halo-rendering approach, i.e. linking the galaxy stellar masses to the underlying mass distribution, with the external shear measurements of the strong lens system. Their combined approach yielded tighter constraints on the inferred external convergence compared to a halo-rendering approach only.

\section{Cosmographic inference} \label{sec7:cosmo_inferences}
Having established the necessary observations and analyses components in the previous sections, in this section we discuss how an end-to-end combined analysis leads to constraints on $H_0$ and other relevant cosmological parameters. First we discuss the analysis for a single lens (Section~\ref{sec7:single_lens}) and then state the analysis for a set of multiple lenses (Section~\ref{sec7:population_inference}).

\subsection{Single lens cosmography}\label{sec7:single_lens}
For each individual strong lens, there are preferrably four data sets available: (i) imaging data of the strong lensing features and the deflector galaxy, $\mathcal{D}_{\rm img}$; (2) time-delay measurements between the multiple images, $\mathcal{D}_{\rm td}$; (3) stellar kinematics measurement of the main deflector galaxy, $\mathcal{D}_{\rm spec}$; (4) line-of-sight galaxy count and weak lensing statistics, $\mathcal{D}_{\rm los}$.
These data sets are independent and so are their likelihoods in a joint cosmographic inference. Hence, we can write the likelihood of the joint set of the data 
\begin{equation}
    \mathcal{D} =\{\mathcal{D}_{\rm img},  \mathcal{D}_{\rm td}, \mathcal{D}_{\rm spec}, \mathcal{D}_{\rm los}\}
\end{equation}
given the cosmographic parameters $\{D_{\rm d}, D_{\rm s}, D_{\rm ds} \} \equiv D_{\rm d, s, ds}$ as
\begin{multline}
    \label{eqn:single_lens_inference}
    \mathcal{L}(\mathcal{D}| D_{\rm d, s, ds} ) =
     \int \mathcal{L}(\mathcal{D}_{\rm img} | \boldsymbol{\xi}_{\rm mass}, \boldsymbol{\xi}_{\rm light}) \\
     \times \mathcal{L}(\mathcal{D}_{\rm td} | \boldsymbol{\xi}_{\rm mass}, \boldsymbol{\xi}_{\rm light}, \lambda, D_{\Delta t})\\
     \times \mathcal{L}(\mathcal{D}_{\rm spec}| \boldsymbol{\xi}_{\rm mass}, \boldsymbol{\xi}_{\rm light}, \beta_{\rm ani}, \lambda, D_{\rm s}/D_{\rm ds})
      \mathcal{L}(\mathcal{D}_{\rm los}| \kappa_{\rm ext}) \\
      \times p(\boldsymbol{\xi}_{\rm mass}, \boldsymbol{\xi}_{\rm light}, \lambda_{\rm int}, \kappa_{\rm ext}, \beta_{\rm ani}) d\boldsymbol{\xi}_{\rm mass} d\boldsymbol{\xi}_{\rm light} d\lambda_{\rm int} d\kappa_{\rm ext} d\beta_{\rm ani}.
\end{multline}
In the expression above we only included the relevant model components of the individual likelihoods. $\boldsymbol{\xi}_{\rm light}$ formally includes the source and lens light surface brightness\footnote{To evaluate the time-delay likelihood, we require the time-variable image positions from the set of $\boldsymbol{\xi}_{\rm light}$ parameters.}.

\begin{figure}[htbp]
\centering
    \includegraphics[width=0.49\textwidth]{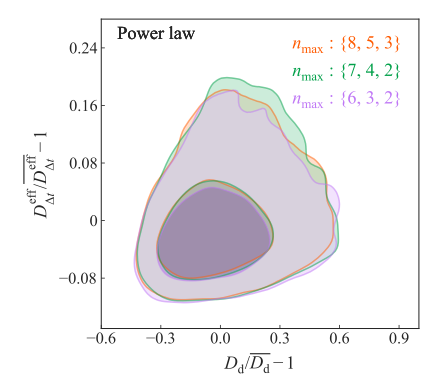}
    \caption{Median-subtracted and mean-divided relative angular diameter distance posteriors, $D_{\Delta t}$ and $D_{\rm d}$ for a power-law mass density profile with three different source reconstruction settings. $n_{\rm max}$ in this figure refers to the polynomial order in the shapelets with the set of three numbers corresponding to three different sources being simultaneously modelled in the image. Figure adopted from \cite{Shajib:2020}.}
    \label{fig:distance posteriors}
\end{figure}

The sampling of the cosmographic posterior from the joint likelihood of Equation~\ref{eqn:single_lens_inference} can be split in parts to simplify the problem. For example, we can first perform the imaging analysis providing constraints on $\boldsymbol{\xi}_{\rm lens}$ and $\boldsymbol{\xi}_{\rm light}$ without sampling the cosmological or distance parameters. In turn, simple sampling the  $\boldsymbol{\xi}_{\rm lens}$ and $\boldsymbol{\xi}_{\rm light}$ posteriors in post-processing when evaluating the time-delay likelihood and stellar kinematic likelihood can translate the posteriors into distance posteriors in $D_{\Delta t}-D_{\rm d}$ space.
Marginalization over different modeling choices can also be done in the $D_{\Delta t}-D_{\rm d}$ posterior space.
Figure~\ref{fig:distance posteriors} provides an example of $D_{\Delta t}-D_{\rm d}$ for a set of different modeling choices.

Weighting the posteriors of different models can be done with Bayesian model comparison. This weighting also allows one to combine models in a single posterior, which then includes systematics considerations. Discrete and finite choices made in the models and scatter in the sampling and BIC calculation can lead to over-constraint model selections. Procedures to take noise and finite model selection in the BIC estimate into account have been developed \citep{Birrer:2019}.

\subsection{Population level analysis} \label{sec7:population_inference}

The overarching goal of time-delay cosmography is to provide a robust inference of cosmological parameters, $\boldsymbol{\pi}$, and in particular the absolute distance scale, the Hubble constant $H_0$, and possibly other parameters describing the expansion history of the Universe (such as $\Omega_{\Lambda}$ or $\Omega_{\rm m}$), from a sample of gravitational lenses with measured time delays.

In Bayesian language, we want to calculate the probability of the cosmological parameters, $\boldsymbol{\pi}$, given the strong lensing data set, $p(\boldsymbol{\pi} | \{\mathcal{D}_{i} \}_{N})$, where $\mathcal{D}_i$ is the data set of an individual lens (including imaging data, time-delay measurements, kinematic observations and line-of-sight galaxy properties) and $N$ the total number of lenses in the sample.

In addition to $\boldsymbol{\pi}$, we denote $\boldsymbol{\xi}$ all the model parameters part of either a single lens analysis (Section~\ref{sec7:single_lens}) or present on the population level. Using Bayes rule and considering that the data of each individual lens $\mathcal{D}_{i}$ is independent, we can write, following \cite{Birrer:tdcosmoiv}:

\begin{multline} \label{eqn:full_inference}
    p(\boldsymbol{\pi} \mid \{\mathcal{D}_{i} \}_{N}) \propto \mathcal{L}(\{\mathcal{D}_{i} \}_{N}| \boldsymbol{\pi}) p(\boldsymbol{\pi})  = \int \mathcal{L}(\{\mathcal{D}_{i} \}_{N} \mid \boldsymbol{\pi}, \boldsymbol{\xi})p(\boldsymbol{\pi}, \boldsymbol{\xi}) d \boldsymbol{\xi} \\
    = \int \prod_i^N \mathcal{L}(\mathcal{D}_{i}| \boldsymbol{\pi}, \boldsymbol{\xi})p(\boldsymbol{\pi}, \boldsymbol{\xi}) d \boldsymbol{\xi}.
\end{multline}

In the following, we divide the nuisance parameter, $\boldsymbol{\xi}$, into a subset of parameters that we constrain independently per lens, $\boldsymbol{\xi}_i$, and a set of parameters that require to be sampled across the lens sample population globally, $\boldsymbol{\xi}_{\rm pop}$. The parameters of each individual lens, $\boldsymbol{\xi}_i$, include the lens model, source and lens light surface brightness and any other relevant parameter of the model to predict the data.
Hence, we can express the hierarchical inference (Eqn. \ref{eqn:full_inference}) as
\begin{multline} \label{eqn:full_inference_extended}
    p(\boldsymbol{\pi} \mid \{\mathcal{D}_{i} \}_{N})
     \propto \int \prod_i \left[ \mathcal{L}(\mathcal{D}_i \mid D_{\rm d, s, ds}(\boldsymbol{\pi}), \boldsymbol{\xi}_i,  \boldsymbol{\xi}_{\rm pop})
     p(\boldsymbol{\xi}_i) \right]
     \\ \times
     \frac{p(\boldsymbol{\pi}, \{\boldsymbol{\xi}_i \}_{N}, \boldsymbol{\xi}_{\rm pop})}{\prod_i p(\boldsymbol{\xi}_i)} d \boldsymbol{\xi}_{\{i\}} d \boldsymbol{\xi}_{\rm pop}
\end{multline}

where $\{\boldsymbol{\xi}_i \}_{N} = \{\boldsymbol{\xi}_1, \boldsymbol{\xi}_2, ..., \boldsymbol{\xi}_N \}$ is the set of the parameters applied to the individual lenses and $p( \boldsymbol{\xi}_i)$ are the interim priors on the model parameters in the inference of an individual lens. The cosmological parameters $\boldsymbol{\pi}$ are fully encompassed in the set of angular diameter distances, $\{D_{\rm d}, D_{\rm s}, D_{\rm ds} \} \equiv D_{\rm d, s, ds}$, and thus, instead of stating $\boldsymbol{\pi}$ in Equation~\ref{eqn:full_inference_extended}, we now state $D_{\rm d, s, ds}(\boldsymbol{\pi})$.
Up to this point, no approximation was applied to the full hierarchical expression (Eqn. \ref{eqn:full_inference}).

Key differences among different inferences of $H_0$ from a set of lenses involve, beyond the assumptions on individual lenses, assumptions on the covariant nature and the prior on the population level of the governing hyper-parameters.
For example, \cite{Wong:2020} assumes full independence of the nuisance priors from one lens to another. Formally, within Bayes Theorem, this approach assumes perfect knowledge of the governing population hyper-parameter distribution prior (Eqn.~\ref{eqn:full_inference_extended}). In this approach, the distance posteriors of individual lenses can be interpreted as measurements and the cosmographic analysis can be done in solely operating in the $D_{\Delta t}-D_{\rm d}$ space with a direct independent and easy accessible likelihood description.

\cite{Millon2020a} performed an analysis exploring the difference between two different radial mass density profile families, assuming that either all lenses are of one type or another, effectively treating one modeling choice as a covariant nuisance parameter in their inference while keeping all other priors independent with an assumed population.

\cite{Denzel:2021} is using a free-form approach in the modeling of individual lenses.
The ensemble of models allowed by the data for an individual lens is providing the model posterior distribution. The underlying regularization scheme is the implicit prior applied on individual lenses. The identical regularization scheme is applied to all lenses assuming independence in the priors without covariances in the choice of the regularization scheme between lenses.
\cite{Denzel:2021} did not use any external information to break the MST. Hence, the specific choice of the regularization scheme with their underlying physical and regularization priors is responsible for the breaking of the MST on the population level.

\cite{Birrer:tdcosmoiv} introduced the hierarchical analysis framework into time-delay cosmography and identified few key parameters, that on a per lens basis are not sufficiently well constrained and thus the population prior can significantly affect the outcome of the analysis. The parameters hierarchically sampled, beyond the cosmological ones, were the MST population $\lambda_{\rm int}$ (Eqn.~\ref{eqn:lambda_combined}), and the stellar anisotropy distribution (see Section~\ref{sec7:external_data}).

\cite{Park:2022} implemented a Bayesian hierarchical framework to determine the external convergence distribution on the population level for a full sample of lens systems used for time-delay cosmography and demonstrated how to correct for a selection bias in the population of lenses when there is limited information on an individual lens basis.

The required population-level description of priors, in particular of parameters that can not be constrained to high precision (overcoming the prior in the analysis) do also need to take accurately into account potential differences among subsets of the population. For example, different lens discovery channels might preferentially select a different lens and line-of-sight population.

%%%%%%%%

\section{Cosmography with galaxy clusters}\label{secX:clusteres}
In this section, we discuss current and past application of cosmograpy with galaxy clusters. We first discuss relative expansion history constraints from multiple source redshifts (Section~\ref{sec:cluster_relative}) and then discuss time-delay cosmography applications (Section~\ref{sec:cluster_h0}). This section is aimed to provide a brief overview over these aspects and we refer to the specific literature referenced in this section for further details.

\subsection{Relative expansion history with galaxy clusters}\label{sec:cluster_relative}

For a given deflector, changing the source redshift alters the angular diameter distances in Equation \ref{eqn:sigma_crit}, whilst the other terms in Equations \ref{eqn:lens_equation}-\ref{eqn:kappa_sigma} are unchanged. Hence, for two photons passing through the same point in the lens plane, but originating on different source planes, the ratio of scaled deflection angles, $\alpha_1$, $\alpha_2$ is given by the cosmological scaling factor, $\beta$,

\begin{equation} \label{eqn:dspl_beta}
    \frac{\alpha_1}{\alpha_2} = \frac{D_{ls1} D_{s2}}{D_{s1}D_{ls2}} \equiv \beta.
\end{equation}

It has been realized that lenses with multiple source planes can additionally provide constraints on cosmological distance ratios sensitive to the relative expansion history and geometry of the Universe \citep[e.g.,][]{Paczynski:1981, Link:1998, Cooray:1999, Golse:2002, Sereno:2002, Soucail:2004, Gavazzi:2008, GilmoreNatarayan:2009}.
\cite{Link:1998} showed that the cosmological sensitivity of the angular size-redshift relation could be exploited using sources at distinct redshifts and developed a methodology to simultaneously invert the lens and derive cosmological constraints.

In particular, galaxy clusters with a large strong lensing cross section do have multiple sources at different redshifts and are exquisite objects to study the geometrical effect between different source redshifts.
In fact, early studies of galaxy clusters already indicated the presence of a cosmological constant \citep[e.g.,][]{Paczynski:1981, Sereno:2002, Soucail:2004}.

The method to probe relative angular diameter distances, in addition to multiple sources at different redshifts, also requires a complete understanding of the lens density profile and other perturbing masses along the line of sight. 

With more exquisite deep multi-colour imaging and spectroscopy for a small subset of galaxy clusters, such as with the Hubble Frontier Fields program \citep{Lotz:2017} has led to the discovery of hundreds of multiple images and thus to a significant improvement of cluster mass estimates \citep{Jauzac:2014, Diego:2016modeling, Lagattuta:2017, Monna:2017}. 
Cluster lenses are complex and most of the efforts have been spent in accurately reconstruct cluster lensing profiles \citep[see e.g.,][]{Jullo:2010, DAloisio:2011, Magana:2015, Caminha:2016, Acebron:2017}.
The mass modelling of strong lensing clusters can be carried out in different manners: parametric and non-parametric methods are equally used; the primary distinction between them being that parametric modelling assumes that luminous cluster galaxies trace the cluster mass whereas non-parametric does not.

The method of using sources at multiple redshifts can also be applied to galaxy-scale lenses, thought double source plane lenses are more rare \cite{Biesiada:2006, Gavazzi:2008, Collett:2012}. The Einstein radius is a function of the lens mass and the cosmological distances. The ratio of Einstein radii in a lens with sources at two or more redshifts is independent of the deflector mass \citep[e.g.,][]{Gavazzi:2008, Collett:2012}. 

In both cases, galaxy and cluster scales, the method also requires a complete understanding of the lens density profile and additional lensing by the source galaxies and other perturbing masses along the line of sight.

\subsection{$H_0$ with galaxy clusters}\label{sec:cluster_h0}

To date, galaxy-scale lenses have dominated the literature on $H_0$ determination in the number of measurements and precision. We have recently witnessed competitive constraints from galaxy clusters in measuring $H_0$ \citep{Kelly:2023, Napier2023, Liu2023, Pascale:2024} and other cosmological quantities \citep{Grillo:2024}.
Massive clusters are rich in multiple images and have definite advantages over individual galaxies. The main one is that sources at multiple redshifts break the mass-sheet or steepness degeneracy \citep[e.g.,][]{Bradac:2004}, which is the main degeneracy and hence source of uncertainty affecting galaxy-scale determination of $H_0$ (see also Section~\ref{sec7:lens_potential}). Having a much larger image separation in clusters compared to galaxy-scale lenses resulting in overall longer time delays of order months to years. Those time delays are relatively easily determined to a few percent precision, rivalling time-delay determinations from quasar sources in galaxy-scale lenses. However, the longer time-delay implies years of monitoring to obtain lightcurves with sufficient overlap in the case of lensed quasars \citep[e.g.]{Fohlmeister2008,Fohlmeister2013,Munoz2022} or dedicated HST follow-up at the time of reappearance in the case of lensed supernovae \citep{Kelly2023b}. 
The drawback of clusters is that their mass distributions are more complex:  they are dynamically younger than galaxies, and their multiple image regions sample a much larger fraction of the clusters' virial radius than in galaxies. Therefore the multiple image region of clusters is expected to be more abundant in substructure, and hence harder to model. These difficulties can be circumvented if there are a few tens or hundreds of multiple images, then $H_0$ can be estimated to a 1-few \% precision \citep{Ghosh2020}. At present, in a cluster lens like MACS 1149, one can estimate $H_0$ to 6\%, assuming a conservative 3\% uncertainty on the observed time delay \citep{Grillo2018}.

The first cluster lens to produce a precise estimate of $H_0$ was MACS 1149, where the first confirmed multiply imaged supernova was observed a few years ago \citep{Kelly2015}. The long time delay before the reappearance of the last arriving image---saddle in the arrival time surface of the cluster---allowed the lensing community to make model predictions for the time of the reappearance. Most models agreed reasonably well on 250-350 day delay \citep{Kelly2016, Treu2016}. Very recently, \cite{Frye:2024} discovered another lensed supernovae in a cluster, this time of type Ia, which even allows for a standardization of the lensing magnification \citep{Pierel:2024}, and \citep{Pierel:encore} discovered for the first time a second supernovae in the same host galaxy, with the initial supernovae discovered in archival data by \citep{Rodney:2021}, further demonstrating the future prospects using cluster-scale lenses.

\begin{figure*}[htbp]
\centering
    \includegraphics[width=0.8\textwidth]{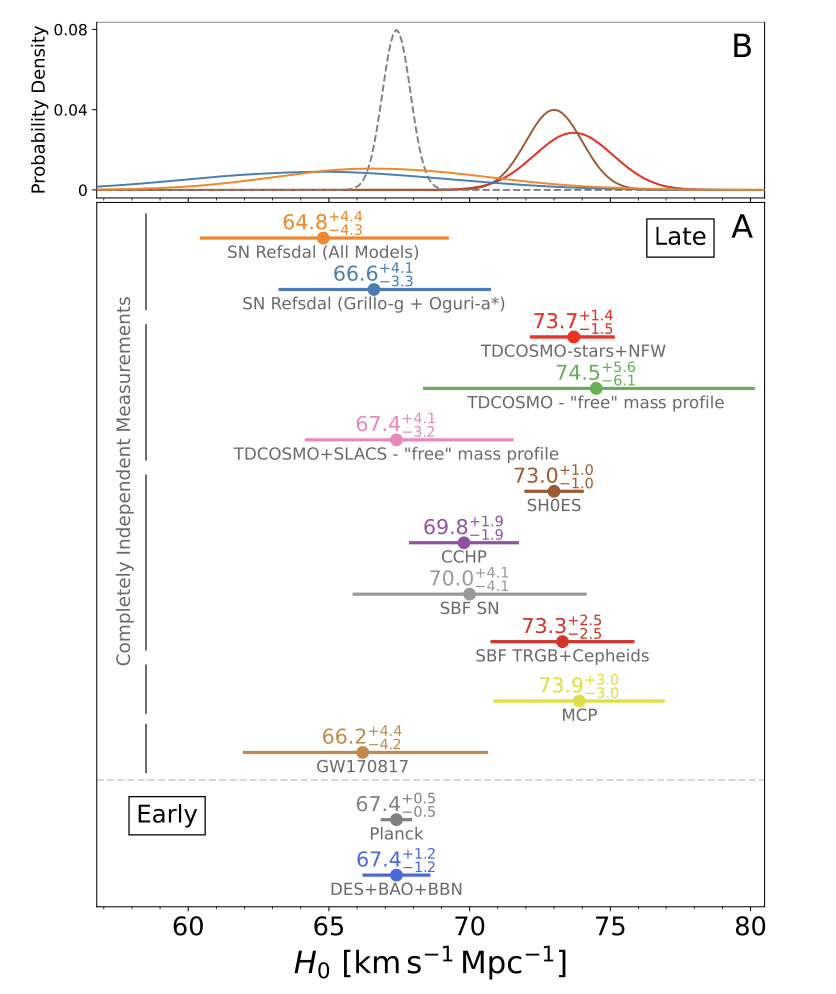}
    \caption{Comparison of recent $H_0$ measurements in the literature. Presented are the time-delay cosmography constraints from \cite{Kelly:2023} of SN Refsdal, for the full set of eight models (orange) and the subset of the two best models (blue), from
    the TDCOSMO collaboration (red)
    of six TDCOSMO time-delay lenses (five H0LiCOW lenses \citep{Wong:2020} and one STRIDES lens by \cite{Shajib:2020} assuming parameteric forms of the mass density profile of the deflector, either described as a power-law or stars (constant mass-to-light ratio) plus dark matter halos \citep{Millon2020a}, from the TDCOSMO collaboration of the same lenses as in \cite{Millon2020a} with virtually no assumption on the radial mass density profile of the lens galaxy, and taken into account the covariance between the lenses (green)
    \citep{Birrer:tdcosmoiv}. The TDCOSMO+SLACS measurement comes from the joint analysis of the TDCOSMO sample and 33 SLACS lenses with SDSS spectroscopy.
    The "free" mass profile assumptions of the two measurements by \cite{Birrer:tdcosmoiv} are constrained only by the stellar kinematics and fully accounts for the uncertainty related to the mass sheet transformation (MST).
    Aside from time-delay studies, the local measurements by SH0ES + Gaia \citep{Riess:2022}, the Carnegie-Chicago Hubble Program (CCHP) \citep{Freedman:2019}, surface brightness fluctuations (SBF) SN \citep{Khetan:2021}, SBF Tip of the Red Giant Branch (TRGB)+Cepheids \citep{Blakeslee:2021}, Megamaser Cosmology Project (MCP) \citep{Pesce:2020}, gravitational wave (GW) event 170817 \citep{Dietrich:2020}, Planck \citep[][; dashed grey]{Planck2020}, and Dark Energy Survey (DES) + Baryon Acoustic Oscillation (BAO) + Big Bang nucleosynthesis (BBN) \citep{DESH0:2018} are presented. Error bars in panel (B) show the 16th, 50th, and 84th percentile confidence levels. Dashed horizontal line separates measurements from observations of the universe early in its evolution from those late in its evolution. $H_0$ measurements bracketed by different vertical gray bars are entirely independent of each other. Figure from \cite{Kelly:2023} which was generated using a previous comparison \citep{Bonvin:2020}.}
    \label{fig:current_results}
\end{figure*}

\section{Current status and results}\label{sec7:current_results}
We go in length through recent results using lensed quasars and also present the recent results using lensed SNe.
Figure~\ref{fig:current_results} summarizes a selection of current measurements and comparison with other probes.

\subsection{Recent results from quasars}

The independent analysis of six lensed quasar systems \citep{Suyu:2010, Suyu:2013, Wong:2017, Bonvin:2017, Birrer:2019, Chen:2019, Rusu:2020} by the H0LiCOW collaboration \citep{Suyu:2017} inferred a Hubble constant value of $H_0 = 73.3^{+1.7}_{-1.8}$  \Hunit. This measurement uses parameteric forms of the mass density profile of the deflector, either described as a power-law or stars (constant mass-to-light ratio) plus dark matter halos following an NFW \citep{NFW} profile with priors on the mass and concentration of the halo reflecting the population of haloes in N-body simulations \citep{Wong:2020}.
The H0LiCOW result is a 2\% precision measurement on H$_0$, in excellent agreement with the local distance ladder measurement by the SH0ES team \citep{Riess:2019, Riess:2021}. Moreover, the H0LiCOW measurement is not more than 3$\sigma$ statistical tension with early-Universe probes \citep[e.g.,][]{Planck2020, Aiola:2020}. 
An additional lens analyzed by the STRIDES collaboration with the same mass profile assumptions as the H0LiCOW collaboration further provided the most precise 
single-lens measurement of $H_0 = 74.2^{+2.7}_{-3.0}$  \Hunit \citep{Shajib:2020}.
In summary, if the mass density profiles of the H0LiCOW and STRIDES lenses are well described by a power-law or a baryonic component with a constant mass-to-light ratio plus dark matter profiles from standard N-body dark matter only simulations, and under the assumption that the covariances are negligible, the tension is significant from the strong lensing measurements alone, and corroborating other measurements \citep[e.g.,][]{Riess:2021}.

Given that incompatibilities between the local value of $H_0$ and the $\Lambda$CDM model extrapolated $H_0$ inference from the CMB or other early-universe physics anchored inherently break the standard model of cosmology and likely may require new physics, several groups, including H0LiCOW, STRIDES and SHARP, now TDCOSMO collaboration, are investigating potential systematics in the $H_0$ measurements.

The TDCOSMO collaboration found, combining six lenses from H0LiCOW, SHARP and STRIDES, that the results when assuming that all lenses are either of one or the other previously assumed forms of the mass density profile are in good agreement with each other when measuring $H_0$. The good agreement in the $H_0$ results between power-law and composite profiles was interpreted by \cite{Millon2020a} as a consequence of the `bulge-halo conspiracy' that the combined baryonic and dark matter density components form a power-law profile \citep[e.g.,][]{Koopmans:2006,Koopmans:2009, vdVen:2009}.
\cite{Denzel:2021} analyzed 8 quadruply imaged quasars with a free-form modeling approach and obtained $H_0 = 71.8^{+3.9}_{-3.3}$ \Hunit.
\cite{Gilman:2020} investigated the effect of unaccounted for subhalos and small undetected line-of-sight halos in the uncertainty budget and found insignificant residual uncertainties to mitigate the tension of the measurements with the CMB and large scale structure probes.
\cite{vdVyvere:2022, vandeVyvere:2022b} showed that a variety of expected azimuthal structures in the mass distribution (i.e. multipoles, twists and ellipticity gradients) should leave $H_0$ unaffected at the population level unless there are specific selection effect in the galaxy population.

The attention further turned to assessing and relaxing the radial profile assumption (see Section~\ref{sec7:mass_profile}), as well as the introduction of population priors for parameters that cannot be constrained on a lens-by-lens basis for a covariant treatment of their uncertainties. \cite{Birrer:tdcosmoiv} addressed the radial profile assumption by choosing a parametrization of the radial mass density profile that is maximally degenerate with H$_0$, via the MST. This is the most explicit and direct way addressing the MST effect on the time-delay cosmographic analysis. With this more flexible parametrization, H$_0$ is only constrained if the measured time delays and imaging data are supplemented by stellar kinematics. Applying this extremely conservative choice to the TDCOSMO sample of 7 lenses increases the uncertainty on H$_0$ from 2\% to 8\% resulting in $H_0 = 74.5^{+5.6}_{-6.1}$  \Hunit, without changing the mean inferred value significantly.

\cite{Birrer:tdcosmoiv} further introduced a hierarchical framework (see also Section~\ref{sec7:population_inference}) in which external datasets can be combined with the time-delay lenses to improve the precision, in particular on the MST parameter of the population, and hence on $H_0$. A secondary required parameter that must be constrained when using stellar kinematics is the stellar anisotropy, due to the mass-anisotropy degeneracy. External data sets with spatially resolved kinematics measurements can aid breaking this degeneracy to constrain the MST parameter.
\cite{Birrer:tdcosmoiv} achieved a 5\% precision measurement on $H_0$ by combining the TDCOSMO lenses with imaging modeling and stellar kinematic measurements of a sample of lenses from the Sloan Lens ACS (SLACS) survey with no time-delay information \citep{Bolton:2008, Auger:2009, Shajib:2021slacs} and measured $H_0 = 67.4^{+4.1}_{-3.2}$ \Hunit. The mean of the TDCOSMO+SLACS measurement is offset with respect to the TDCOSMO-only value, effectively matching the CMB inferred value, although still statistically consistent with previous assumptions given the uncertainties in the measurement. The \cite{Birrer:tdcosmoiv} measurements with and without the SLACS dataset added are in statistical agreement with each other and with the earlier H0LiCOW/ SHARP/ STRIDES measurements based on the radial mass profile assumptions. The result by \cite{Birrer:tdcosmoiv} is also consistent, with the work by \citet{Shajib:2021slacs} studying more flexible mass density profiles and mass-to-light gradients. The studies by \cite{Birrer:tdcosmoiv} and \cite{Shajib:2021slacs} share the same measurements for the SLACS lenses and the consistency is implied by construction. \citet{Shajib:2021slacs} concluded that NFW+stars,  when using wider priors on mass and concentration than earlier H0LiCOW/ SHARP/ STRIDES measurements, is a sufficiently accurate description of the mass density profile of the SLACS lenses. However, a larger flexibility in the  mass-concentration relation on the population level and small departures from those radial forms are allowed by the data, resulting in the uncertainties reflected by the \cite{Birrer:tdcosmoiv} analysis. The shift in the mean in $H_0$  when adding the SLACS lenses could be real or it could be due to an intrinsic difference between the deflector population in the TDCOSMO and SLACS samples. Differences in the deflectors might arise from unequal selection effects. For example, the two samples are well matched in stellar velocity dispersion (total mass), but they differ in redshift. Potentially unaccounted for evolutionary trends in the mass profiles could bias the results when adding samples of lenses at different redshifts. Another example is that the TDCOSMO sample is source selected, meaning the main characteristics for the data set to be discovered and selected are properties of the source as seen when lensed, and composed mostly of quadruply imaged quasars, while the SLACS sample is deflector selected, meaning the primary criteria for the selection arises from properties of the deflector irrespective of the source and its geometric lensing effect, and dominated by doubly imaged galaxies.

More recently, \cite{Shajib:2023} performed the first analysis with spatially resolved stellar kinematics measurement with the same conservative assumptions as \cite{Birrer:tdcosmoiv} and achieved a $\sim 9\%$ measurement from a single quadruply lensed quasar, finding results in agreement with the previous analysis based on power law and composite models \cite{Suyu:2014}. This work demonstrates the constraining power of kinematic data in the absence of priors on the shape of the mass density profile.

\subsection{Recent results from lensed supernovae}
In addition to lensed quasars, the discovery of the first multiply-imaged supernovae (SN) \citep{Kelly:2015} in the cluster MACS 1149 allowed to measure $H_0$ with lensed SNe \citep{Vega-Ferrero:2018, Kelly:2023}.
\cite{Kelly:2023} presents results of a combination of several models from different independent modeling teams which were done truly blind, before the measured time delays were known. A Bayesian model selection was done based on the precision and accuracy of the predicted image positions of the reappearance of SN Refsdal, and the predicted magnification ratio (which are independent of H$_0$). Combining these Bayesian weights with the weighted uncertainties of all the eight individual models, they found $H_0 = 64.8^{+4.4}_{-4.3}$ \Hunit. \cite{Kelly:2023} found that models that assign dark-matter halos to individual galaxies and the overall cluster best reproduce the observations. When combining the two best performing models, both consistent within their uncertainties with each other, \cite{Kelly:2023} found $H_0 = 66.6^{+4.1}_{-3.3}$ \Hunit.
Very recently, a $H_0$ measurement was made by a second lensed SN, supernova "H0pe" \citep{Frye:2024}. A combination of a spectroscopic and photometric time-delay measurement \citep{Chen:2024, Pierel:2024} were compared to the predictions of seven independently constructed cluster lens models to measure a value for the Hubble constant. In combination with the standardizable magnification of the type Ia nature of the SN, this resulted $H_0 = 75.4^{+8.1}_{-5.5}$ \Hunit.
These are very encouraging and precise measurements.
For further discussions of lensed SNe, we refer to \chapsn, a focused chapter about lensed SNe of this series.

%%%%%%%%

\section{Outlook in the (near) future} \label{sec7:outlook}

The goal of time-delay cosmography is to provide a robust measurement of the Hubble constant to 1\% precision to decisively tell the outcome of the currently observed tension between late and early time measurements of $H_0$. In the previous Section~\ref{sec7:current_results} we presented current results.
In this section, we discuss the potential of the time-delay method in the near future. To do so, we first provide some details on the error budget of current analyses which allows us to assess and scale to the expected future samples and data quality (Section~\ref{sec7:error_budget}).
Second, we describe the data and instrumentation which enable us to push ahead (Section~\ref{sec7:future_data}). Third, we will highlight avenues where continuing work is required in assessing the methodology to maintain accuracy while increasing the precision of the measurements (Section~\ref{sec7:future_methods}). Finally, we leave some concluding remarks about the prospects of time-delay cosmography in Section~\ref{sec7:concluding_remarks}.

\subsection{Error budget}\label{sec7:error_budget}

Table~\ref{tab:error_budget} presents an overview of the
error budget of time-delay cosmography divided into different aspects of the analysis for individual lenses, the current work of 7 lenses, and a forecast of a future analysis of 40 lenses with improved data. The error budget is split between the three different components of time-delay cosmography: The time-delay measurement (Section~\ref{sec7:time_delays}), the main deflector profile for the prediction of Fermat potential (Section~\ref{sec7:lens_potential}), and the line-of-sight contribution (Section~\ref{sec7:los}). For the deflector error budget, we provide an uncertainty on the lens model excluding the potential systematics of the MST, which can be achieved with high-resolution imaging, and a total error budget including the MST, which depends on additional data (such as currently stellar kinematics of the deflector). For the analysis of individual lenses, we report uncertainties from \cite{Wong:2020, Millon2020a} excluding the MST, and from \cite{Birrer:tdcosmoiv} including the MST in the deflector profile.
The errors between the time-delay measurement, deflector model and line of sight are effectively uncorrelated, and hence the total error budget adds in quadrature. Excluding the MST-related uncertainties, the three components are roughly on equal footing, and the best single lenses alone can provide an $H_0$ uncertainty of $<4\%$ \citep[e.g.,][]{Shajib:2020}.
With stringent priors on the deflector potential ignoring additional MST-related uncertainties, the current sample of 7 lenses assuming uncorrelated uncertainties result in a $\sim 2\%$ uncertainty on $H_0$ \citep{Wong:2020, Millon2020a}.

The MST-related uncertainty for an individual lens depends on the imposed priors and can be constrained to a $\sim 10-20\%$ level with current pre-JWST kinematic measurements. The combination of 7 lenses by \cite{Birrer:tdcosmoiv} lead to a $\sim 8\% $ uncertainty on $H_0$, dominated by a $\sim 7\%$ uncertainty on the MST. Given the uncertainties and their expected scaling ($\Delta H_0/H_0 \propto \sqrt{N}$ for $N$ number of lenses), for 40 lenses, the MST-related uncertainty is the dominant and single most relevant uncertainty to achieve a 1\% measurement of $H_0$. The required data and methodology improvements are laid out in the following subsections.

The secondary uncertainties of the relative expansion history on $H_0$ is $< 1\%$, not mentioned in Table~\ref{tab:error_budget}, but included in the total error budget on the population and can be mitigated by either an informed prior on the relative expansion history, or by a combination with probes sensitive to the relative expansion history.

\begin{table*}
	\centering
	\small\addtolength{\tabcolsep}{-0pt}
	\small
	\caption{Approximate error budget of time-delay cosmography divided into different aspects of the analysis for individual lenses, the current work by \cite{Birrer:tdcosmoiv} of 7 lenses, and a forecast of a future analysis of 40 lenses with improved data. The error budget is split between the three different components of time-delay cosmography: The time-delay measurement (see Section~\ref{sec7:time_delays}), the main deflector profile for the prediction of Fermat potential (see Section~\ref{sec7:lens_potential}, and the line-of-sight contribution (see Section~\ref{sec7:los}. For the deflector error budget, we provide an uncertainty on the lens model excluding the potential systematics of the MST (ex MST), which can be achieved with high-resolution imaging, and a total error budget including the MST, which depends on additional data (such as currently stellar kinematics of the deflector). The forecast for 40 lenses is based on the same mix of quality in the data as for the current 7 lens constraints.}
	\label{tab:error_budget}
	\begin{threeparttable}
	\begin{tabular}{lrrr} % four columns, alignment for each
		%\hline
         & Individual lens & 7 lens sample & 40 lens sample \\
         \hline
         \hline
        Time delay & $1-20\%$ & $\lesssim 0.5\%$ &  $\lesssim 0.1\%$\\ 
         \hline
        Deflector model (ex MST) & $1 - 10\%$ & $\sim 0.8\%$ & $\lesssim 0.3\%$\\
        % \hline
        Deflector model & $10-20\%$ & 7\% & $\lesssim 1\%$\\
         \hline
        Line of sight & $1-5\%$ & $\lesssim 0.5\%$& $\lesssim 0.2\%$\\
         \hline
         \hline
         Total (ex MST) & $3-20\%$ & $2.3\%$& $\lesssim 0.5\%$\\
         Total & $10-25\%$ & $8\%$& $\sim 1\%$\\
         \hline
         \hline
	\end{tabular}
    \end{threeparttable}
\end{table*}

\subsection{Future data} \label{sec7:future_data}
We expect there to be several 10'000 galaxy-galaxy lenses, several hundred quadruply lensed quasars and more than a thousand doubly lensed quasars on the full sky \citep[e.g.,][]{Oguri:2010, Collett:2015}. With the upcoming large area (wide) and sensitive to faint objects (deep) ground- and space-based surveys, such as the Vera C. Rubin Observatory, Roman Observatory, and Euclid, we expect many of those lenses to be discovered within a decade.
Compared to the current analyses conducted on few lenses (e.g., 7 lenses in case of current TDCOSMO results), these are several e-foldings of the number of lenses possibly suitable for time-delay analyses. The sheer number of lenses  will transform the measurements and new approaches are going to be required in the domain of time-delay cosmography to efficiently and accurately make use of all the data available.

The first step in utilizing these lenses present on the sky is to discover them in the large data sets. We refer to \chapsearch~ for an extensive review of techniques, recent successes and an outlook in the searches and discovery of strong lenses. 
The next step is to acquire all the necessary follow-up data products to conduct accurate and precise cosmographic analyses (see Section~\ref{sec7:analysis_overview} and subsequent sections). The data products range from monitoring data for a time-delay measurements, high-resolution imaging, to spectroscopic information about the source and lens redshift as well as velocity dispersion of the deflector. This step is very resource expensive and there are going to be challenges in how to allocate these limited resources. 
Decisions will have to be made to decide which lenses are followed-up. We comment in Section~\ref{sec7:future_methods} about developments of methodology that can deal with less constraining or incomplete data for a larger lensing data set.
Some lenses might require less substantial monitoring follow-up in case where LSST light curves are good enough for a time-delay measurement \citep{Liao:2015}. Some lenses may also automatically obtain high-resolution and sufficiently high signal-to-noise ratio imaging data from wide field space surveys, such as Euclid or Roman \citep{Meng2015}.
Understanding to what extent the acquired data products impact the precision on $H_0$ is key to assess the need for allocating follow-up resources and on which lenses to spend it.
Besides the limited resources, follow-up decisions are currently also impacted by the limited access to adaptive optics (AO) instrument on ground-based large-diameter telescopes.
With the next-generation AO instrumentation and their commissioning on both hemispheres, we expect a full sky accessibility that allows the community to target every single gravitational lens on the sky.

The dominant uncertainty in the current measurement of the Hubble constant with time-delay cosmography is attributed to uncertainties in the mass profiles of the main deflector galaxies (see e.g., Section~\ref{sec7:mass_profile}).
There are multiple independent avenues available in the near future to approach a 1\% measurement of $H_0$ with different data sets. We will focus on these pathways with improved instrumentation and increased data sets in this section.

\begin{figure*}[htbp]
\centering
    \includegraphics[width=0.9\textwidth]{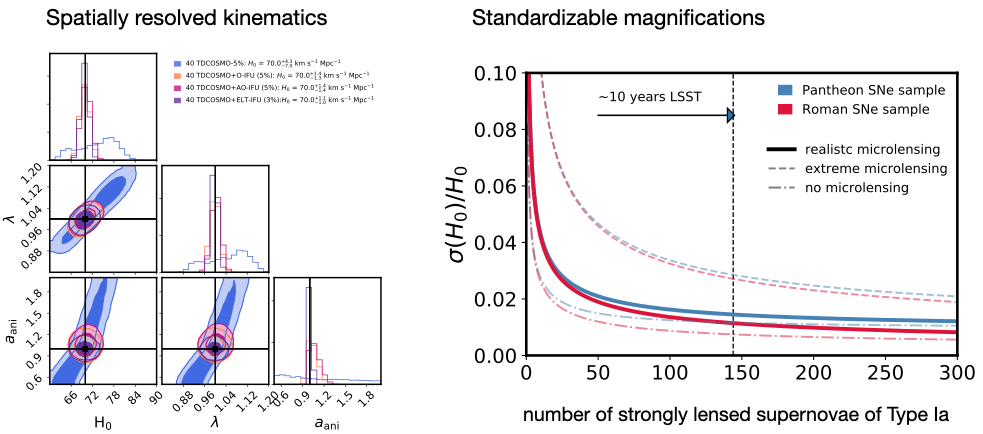}
    \caption{Forecast for $H_0$ measurements in the near future with the upcoming facilities. Left: Spatially resolved kinematics measurements of a sample of 40 time-delay lenses enable a precision on $H_0$ of 1.5\% JWST (Figure adopted from \cite{BirrerTreu:2021}). Right: Lensed supernovae with standardizable magnification measurements. An expected yield  of $\sim 144$ gravitationally lensed supernovae over the span of the 10 years LSST survey enable a precision on $H_0$ of 1.5\% (Figure adopted from \cite{Birrer:2021glSNe}). Both approaches, stellar kinematics and standardized magnifications, do provide independent observational constrains on the MST with different systematics.}
    \label{fig:forecast}
\end{figure*}

Spatially resolved stellar kinematics of the deflector galaxy (see Section~\ref{sec7:external_data} for details on methodology) with the next generation space (James Webb Space Telescope; JWST) and ground-based (extremely large telescopes; E-ELT, GMT, TMT) instruments provide precise measurements of the kinematics of stars. Such two-dimensional observations of the kinematics, paired with the lensing measurements, have the ability to break the mass-anisotropy degeneracy, a currently limiting systematic when interpreting and de-projecting integrated kinematic measurements to measure the three-dimensional gravitational potential. \cite{BirrerTreu:2021} forecasts, based on the methods and assumptions used by \cite{Birrer:tdcosmoiv} without relying on mass-density profile assumptions to break the MST, that with 40 time-delay lenses with exquisite spatially resolved kinematics and otherwise similar measurements as the 7 lens TDCOSMO sample, a 1.5\% precision on $H_0$ can be achieved (Figure~\ref{fig:forecast} left graphic), and see also e.g. \cite{Yildirim:2020, Yildirim:2021}. Such a strategy with exquisite data on the sample of time-delay lenses is one way to make progress. Another approach is to infer the mass density profile properties from a larger set of non-time-delay lenses and apply the constraints on the mass density profile and stellar anisotropy distribution on the time-delay lenses \citep{Birrer:tdcosmoiv,  BirrerTreu:2021, Gomer2022}. In particular, resolved spectroscopy can also be employed on non-time delay lenses without bright and contaminating quasar images, either as prior constraints or by directly incorporating into a hierarchical analysis, to further improve the kinematic measurement precision.

Standardizable magnifications with gravitationally lensed supernovae (glSNe) provide another promising avenue to constrain the mass density profiles and open up an avenue for a percent measurement of $H_0$ in the near future with the onset of LSST. Standardizable magnifications are able to constrain the absolute lensing magnification and hence constrain the density profile (incl. the MST) \citep[e.g.,][]{Kolatt:1998, Oguri:2003, Foxley-Marrable:2018, Birrer:2021glSNe}.
\cite{Birrer:2021glSNe} (Figure~\ref{fig:forecast} right panel) provides a forecast with glSNe in constraining $H_0$ independently of stellar kinematics. They conclude that the standardizable nature of glSNe of type Ia enables a 1.5\% $H_0$ measurement with a 10 years LSST survey. This forecast is contingent to a near-optimal discovery and follow-up effort of glSNe. We refer to \chapsn~ for a detailed review and in-depth discussion on the discovery, expected number of glSNe, the challenges of following them up and the caveats of micro-lensing.

Another method is to make use of the statistical distribution of images under the assumption of knowing the distribution of sources in the source plane with a statistical combination of a large sample of time-delay lenses, relying purely on strong lensing data \citep{Sonnenfeld:2021}.

Yet another method to constrain the radial density profile is to use galaxy-galaxy weak gravitational lensing for a large sample of deflectors analogue to the strong gravitational lenses \citep{Khadka:2024}.

Overall, the trade-offs of analysing all (or most) of the lenses, with most of them with limited data, or to focus on a few of the best (e.g. "golden") lenses, has yet to be seen and explored in detailed. Different approaches have advantages and inconveniences in regard to precision and accuracy on the $H_0$ measurements.

\subsection{Methodology improvements} \label{sec7:future_methods}

With the expected wealth of data and the increase in the number of time-delay and non-time-delay lenses, the prospect of measuring $H_0$ to 1\% precision can become a reality. The employed methodology and assumptions must keep up to provide the accuracy requirement.
In the following we discuss methodology improvements and validations in the domain of galaxy density profiles (Section~\ref{sec7:future_methods_mass_profile}), assumption in the interpretation of non-lensing constraints (Section~\ref{sec7:future_methods_non_lensing}), selection effects (Section~\ref{sec7:future_methods_selection_effects}), automatization (Section~\ref{sec7:future_methods_automatization}) and general aspects of methodology verification (Section~\ref{sec7:future_methods_verification}).
These sections are not meant to be complete but to provide guidance in the near future on where focused effort is required.

\subsubsection{Galaxy density profiles}\label{sec7:future_methods_mass_profile}

The currently employed model mitigating the MST effect by \cite{Birrer:tdcosmoiv} is parameterized with a pure MST parameter $\lambda$. This parameterization is foremost of mathematical nature and leaves the physical interpretation \citep[e.g.,][]{Blum:2020} ambiguous. A pure MST parameterization may in certain regimes even become unphysical, e.g. resulting in total mass profiles with negative density in the outskirts\footnote{Projected lensing convergence can come slightly negative to the extent that the physical mass drops below the mean background density.}.
Such a one-parameter extension to previously considered more simple and rigid mass profiles may also not encompass the necessary flexibility beyond the pure MST that can affect kinematics observations \citep[e.g.,][]{Birrer:tdcosmoiv, Yildirim:2021}, or to deal with more generalized forms of lensing degeneracies, such as the SPT.
To make progress, the full degeneracy inherent in gravitational lensing needs to be folded into flexible, but physically motivated, mass profile parameters. Such an approach was explored by \citep{Shajib:2021slacs} constraining the extended mass density profiles of the SLACS galaxy-galaxy lenses, but has not yet been employed for time-delay cosmography.
Quasar microlensing studies might also help to constrain the stellar mass to light ratio in massive elliptical galaxies. Ambitious measurements below the 10\% level might additionally help to constrain the mass density profiles and would allow the focus on the dark matter portion of the profile. We refer to \chapmicro~ for techniques and prospects of this methodology.

\subsubsection{Non-lensing constraints} \label{sec7:future_methods_non_lensing}
Significant constraints on the MST, and in general mass density profiles, are expected to come from non-lensing observables. These measurements, as well as their model interpretation, need to be tested to the percent level.
For example, for the kinematic measurements, the impact of stellar template fitting needs to be further assessed and validated. For the interpretation of the measurements, de-projection assumptions, rotational structure, as well as stellar anisotropy need to be rigorously tested and assessed for covariant systematics on the population level.
For upcoming magnification measurements with glSNe, micro- and milli-lensing effects need to be assessed and incorporated into the model self-consistently. Furthermore, more knowledge about the structure and size of the variable quasar accretion disc are required to determine the strength of the micro-lensing time delay effect.

\subsubsection{Selection effects} \label{sec7:future_methods_selection_effects}

The phenomena of strong gravitational lensing is inherently a very specific selection effect of an otherwise mostly weak lensing field. 
Quantifying the selection effect of where and in what form strong lensing phenomena occur is going to be crucial requirement to maintain accuracy when increasing precision on $H_0$ in the years to come.
The strong lensing phenomena is impacted by both, the nature of line-of-sight structure, and the main deflector. For the line-of-sight structure, the convergence either raises or lowers the lensing efficiency of an equal mass galaxy to act as a strong lensing deflector, and the cosmic shear changes the geometry of the caustic structure, making it more likely to have quadruply imaged sources.
Similarly, for the main deflector, more concentrated mass distribution, or favorable projections along the line of sight, lead to higher lensing efficiencies, and more elliptical mass profiles (also in projection) lead to a more extended inner caustic region.
Including the differential selection effects among different samples of lenses is required when combining information coming from differently selected populations.
For example, quadruply lensed quasars are visible only when the source quasar lies within the diamond caustic of the lensing galaxy. This condition creates a Malmquist-like selection effect in the population of observed quadruply lensed quasars, increasing the true caustic area \citep{Baldwin:2021}.

Many of these effects are hard or near-impossible to quantify on a lens-by-lens basis. These selection effects need to be modeled and inferred on the population level, with the focus of making sure that relative selection effects between different sub-populations are being understood.

There are two distinct and complementary approaches to understand and mitigate selection effects in the analysis. First, one can attempt to understand the selection from first principles which then can be used to explicitly account for in the analysis procedure. This approach requires extensive simulations including all relevant aspects, starting from the full sky a prior abundance and population of phenomena and a reproducible selection function of the discovery channel and follow-up decision being made to either include a lens in the sample or not.
For example, \cite{Collett:2016} simulated a sample of double- and quadruple-image systems and when assuming reasonable thresholds on image separation and flux, based on current lens monitoring campaigns, they found that the typical density profile slopes of monitorable lenses are significantly shallower than the input ensemble.
Second, one can empirically determine a relative selection function by comparing a set of observables of a sample of lenses compared to random galaxies or sight lines on the sky. Deviations on the set of observables on the population level indicate then the level of selection bias in the sample. Observables may include, but are not restricted to, central velocity dispersion, stellar mass, size and morphology of the deflector, number of subhaloes and line-of-sight projected galaxies nearby, redshift of the deflector, among others. Deviations from established scaling relations among the galaxy properties are then indications of selection biases
We refer to Section~\ref{sec7:los} for data and approaches to quantify line of sight effects. We also stress that these techniques rely on underlying priors and model assumption on the population bias and an explicit de-biasing is required to constrain hierarchically unknown selection effects \citep[see e.g.,][]{Park:2022}.
Currently, neither of the two approaches have been successfully applied.

With the expected large number of lenses in the next few year of the mid 2020s, and the more uniform data set of large and deep surveys, both, the theoretical forward modeling and the empirical hierarchical modeling, will become feasible. We also advocate for analyses to take into account the specific discovery channel of the lenses when performing population level inferences. Understanding the selection function may or may not imply to effectively re-discover the lenses in the analysis to guarantee a uniform and reproducible selection and analysis.

\subsubsection{Automatization} \label{sec7:future_methods_automatization}

Current state-of-the-art analyses of single lens systems takes up more than a year of work, with the involvement of many people, as well as several hundred of thousands of CPU hours of computational cost. To utilize the upcoming larger lens samples and to achieve a high-precision $H_0$ measurement, the time to analyse a single system has to be reduced significantly. Automated decision-making and model choices \citep[e.g.,][]{Schmidt23, Ertl23}, as well as GPU assisted computations \citep[e.g.,][]{Gu:2022} hold promises in these regards.
Moreover, analyses have to be able to be repeated with modifications to test for assumptions covariant among all lenses multiple times. The faster the entire analysis runs, the more explorations of potential systematics in the choices can be executed. The challenge in finding uniform analyses choices are that every lens is different from another and particularities have been notices that needed special attention for lenses on the individual basis.
The analyses conducted need to be uniform in their choices and approaches such that impacts on assumptions can be tested on the ensemble level. Uniformity of analyses can also reduce human errors and sets the analyses on quantifiable priors.

There is currently an effort in homogenizing the analysis procedure, for both time-delay lenses \citep{Shajib:2019, Schmidt23, Ertl23} and non-time delay lenses \citep{Shajib:2021slacs} and further effort is underway.
In parallel, alternative methodology in the modeling and posterior inference are being explored with machine learning techniques, which have the potential to speed up the analysis by orders of magnitude \citep[e.g.,][]{Park:2021}

\subsubsection{Methodology verification} \label{sec7:future_methods_verification}

Guaranteeing accuracy with ever more precise measurements is a challenge throughout the cosmological community. High-precision measurements of quantities to relevance of fundamental physics is a relatively new field and we dedicate a separate subsection highlighting different strategies to verify the methodology and to perform to the necessary quality standard to maintain accuracy.

\begin{itemize}
    \item Realistic simulations offer a validation of a methodology on a known truth \citep[see e.g.,][]{Xu:2016, Tagore:2018}. It is important that the complexity in the simulations are realistic to explore avenues of potential systematics and gain a deep understanding of what data products are able to constrain what aspects of the model. Simulations eventually need to encompass all aspects of the analysis, including the selection effect and the entire line-of-sight structure within the full cosmological and astrophysical context.

    \item Data modeling challenges, such as the Time-Delay Challenge \citep{Liao2015} and the Time-Delay Lens Modeling Challenge \citep{TDLMC} offer platforms to validate currently employed methodology on mock data sets, explore new ways of analyzing the data and can provide a transparent overview of the current state of the field.
    
    \item Blind analyses prevent experimenter bias. The analysis should be guided by the assessment of uncertainties regardless of the anticipated result. Blind analyses have regularly been performed by the H0LiCOW and TDCOSMO collaborations.
    
    \item Open source accessibility of the raw data, processed data product, analysis software and entire end-to-end analysis pipelines can best guarantee reproducibility, form community trust and provides access to the community to alter and improve existing methodology. 
\end{itemize}

\subsection{Concluding remarks} \label{sec7:concluding_remarks}

Time-delay cosmography has an exciting time ahead.
The method has come along way since its original proposal by \cite{Refsdal:1964}. Current measurements of the Hubble constant with time-delay cosmography are at the few percent level, enabled by detailed analyses and precise measurements of different aspects of the analysis. With the expected increase in the lensing sample and the advances in instrumentation, the path towards a percent precision measurement of $H_0$ becomes in reach.

Measuring the Hubble constant to percent level precision is a challenging endeavor, regardless of the cosmological probe.
In this manuscript, we aimed to provide a detailed account of the methodology and measurements to provide guidance to achieve a precise and accurate measurement of $H_0$ at the one-percent level. We emphasized the challenges and systematics in the different components of the analysis and strategies to mitigate them.
Above all, in Carl Sagans words: "Extraordinary claims require extraordinary evidence".

\begin{acknowledgements}
We thank the International Space Science Institute in Bern (ISSI) for their hospitality and the conveners for organizing the stimulating workshop on $``$Strong Gravitational Lensing$"$.
We thank Liliya Williams for contributions to Section \ref{secX:clusteres} and useful comments, and Paul Schechter for useful comments in the preparation of the manuscript.
SB is partially supported by the Kavli Foundation and Stony Brook University.
MM acknowledges support by the SNSF through grant P500PT\_203114 and funding from the European Research Council (ERC) under the European Union's Horizon 2020 research and innovation programme (COSMICLENS : grant agreement No 787886).
AJS is supported by NASA through the NASA Hubble Fellowship grant HST-HF2-51492 by the Space Telescope Science Institute, which is operated by the Association of Universities for Research in Astronomy, Inc., for NASA, under contract NAS5-26555.
FC and DS acknowledge funding from the European Research Council (ERC) under the European Union's Horizon 2020 research and innovation programme (COSMICLENS : grant agreement No 787886). 
SHS thanks the Max Planck Society for support through the Max Planck Research Group.  This research is supported in part by the Excellence Cluster ORIGINS which is funded by the Deutsche Forschungsgemeinschaft (DFG, German Research Foundation) under Germany's Excellence Strategy -- EXC-2094 -- 390783311.
TT acknowledges support by the National Science Foundation through grants AST-1906976 and AST-15436, by NASA through grants HST-GO-15652, and by the Gordon and Betty Moore Foundation through grant 8548.

\end{acknowledgements}

\noindent
{\small \textbf{Conflict of interest} \,
The authors declare no competing interests.}

\bibliographystyle{aps-nameyear}      % American Physical Society (APS) style, author-year citations
\bibliography{references}                % name your BibTeX data base
\nocite{*}

\end{document}